%% file: main.tex
\documentclass[12pt]{article}
\usepackage[margin=1in]{geometry}
\usepackage{graphicx, amsmath, bm, amssymb, cleveref}
\usepackage{comment}
\usepackage[sorting=none, backend=biber]{biblatex}
\usepackage{xcolor}
\usepackage[normalem]{ulem}

\DeclareSourcemap{
  \maps[datatype=bibtex]{
    \map{
      \step[fieldset=urldate, null]
      \step[fieldset=url, null]
    }
  }
}

\title{Thermal Stress Disrupts Symbiotic Fluid Dynamics in Bobtail Squid}
\author{Stephen Williams$^1$, Kyra Alexa Ruiz$^1$,\\ 
Elizabeth Heath-Heckman$^2$, Erica M. Rutter$^1$, Shilpa Khatri$^1$\\
$^1$University of California, Merced, Department of Applied Mathematics\\
$^2$Michigan State University, Department of Integrative Biology \\
$^*$Corresponding Author: swilliams64@ucmerced.edu}

\date{} 

\begin{document}

\maketitle

\input{abstract}

\input{introduction}

\input{methods}

\input{results}

\input{discussion}

\input{other}

\printbibliography

\input{SM}

\end{document}

%% file: abstract.tex
\begin{center}
    Keywords: Biofluids, Method of Regularised Stokeslets, Stokes Flow, \\
    Sensitivity Analysis, Symbiosis, Mutualism, Euprymna scolopes, Vibrio fischeri, Climate Change
\end{center}

\begin{abstract}

The impact of thermal stress on beneficial symbiosis, in the face of rapid climate change, remains poorly understood. 
We investigate this using the model system, \textit{Euprymna scolopes}, the Hawaiian Bobtail Squid, and its bioluminescent symbiont, \textit{Vibrio fischeri}, which enables the squid to camouflage itself through counter-illumination. 
Successful colonisation of the squid by \textit{V. fischeri} must occur hours after hatching and is mediated by fluid flow due to respiration within the squid mantle cavity. 
To study this process, we develop a mathematical model using the Method of Regularised Stokeslets to simulate the flow and resulting bacterial trajectories within the squid.
We explore how thermal stress, mediated by physiological changes in respiration, ciliary dynamics, and internal geometry, affects this early colonisation by analysing the time bacteria spend in regions crucial to the establishment of symbiosis in these simulations. 
A variance-based sensitivity analysis of physiologically relevant parameters on these metrics demonstrates that changes in the breath cycle significantly impact and reduce the time bacteria spend in the critical zone within the squid, hindering colonisation.

\end{abstract}

%% file: introduction.tex
\section{Introduction}

The Hawaiian Bobtail Squid, \textit{Euprymna scolopes} (Es), is a small nocturnal cephalopod endemic to several shallow-water locations around Hawaii.
These locations are home to many marine predators of the squid, particularly the Hawaiian monk seal \cite{curtice_home_2011}. 
As such, Es have developed several strategies to evade predation in their environment, including spending daylight hours buried in the seabed and hiding from predators during the most active hours of the day.
This strategy is not viable at night, when the squid must emerge to hunt for food.
An individual Es in the water column at night will produce a silhouette when viewed from below, due to downwelling moonlight and starlight, allowing nearby predators to locate the squid. 
However, Es has developed a symbiotic relationship with the bioluminescent bacterium, \textit{Vibrio fischeri} (Vf), which allows the squid to evade predation \cite{mcfall-ngai_symbiont_1991}. 

The light from Vf enables Es to perform a process known as counter-illumination, in which their shadow is obscured, protecting them from predators \cite{Jones2004}.
In return, the symbiotic bacteria are provided with nutrients and attain a higher density in the environment than they would otherwise.
To house their bacterial partners, bobtail squid produce an internal structure, the light organ, which maintains an active symbiont population. 
From the light organ, in conjunction with several bodily features (transparency, internal filters, lensing, etc.), mature squid can modulate the quantity and, to some extent, spectrum of light they produce ventrally \cite{montgomery_embryonic_1993}.
The effective and prompt establishment of symbiosis between juvenile Es and Vf is crucial for squid survival \cite{nyholm_lasting_2021,visick_lasting_2021}.
As such, Es has several internal structures to harvest Vf from the surrounding environment and to allow the bacteria to colonise the light organ.
This bacterial selection is primarily facilitated by bilaterally symmetric ciliated appendages that are attached to the juvenile light organ.

\input{figures/figure1}

There are several challenges that the Es-Vf system must overcome for the bacteria to colonise the squid.
The first hurdle a juvenile squid must overcome is access to free-living Vf.
Adult Es vent approximately $90\%$ of their internally stored bacterial population each morning. 
The reasons for this venting are not fully understood. 
One hypothesis is that venting seeds the seawater with cooperative strains of Vf, allowing successive generations of Es to be colonized with appropriate symbionts.
Another hypothesis is that venting enables the dynamic selection of Vf best suited for the squid's needs throughout the life of the host.
Regardless of the reasons for this venting, since adult and juvenile squid tend to live in proximity, the juvenile habitat has a viable bacterial population due to the adult venting.
This proximity, combined with the ubiquity of Vf bacteria across global marine environments, means that newly hatched juveniles have ready access to their potential symbionts.

The next step in acquiring symbionts for a juvenile squid is to bring the bacteria into the mantle cavity from the environment and select for Vf. 
To respire, squid must expand and contract their mantle to bring oxygenated water into their body cavity and over their gills, their main respiratory organs.
As the light organ is located directly anterior to the gills in juvenile Es, respiration results in water flowing directly over the light organ (see \Cref{fig:figure1}A and \ref{fig:figure1}B).
Once the water containing bacteria is brought into the internal cavity of the Es, the squid must identify and select the desired Vf, and discard other unwanted particles back into the environment.
Previous work has begun to disentangle how fluid-structure interaction at the surface of the light organ enables the squid to filter desirable bacteria from the respired fluid \cite{nawroth_motile_2017}. 
In particular, \cite{nawroth_motile_2017} found that beating cilia on the light organ appendages create a directional flow, enabling particles of a particular size to be captured (see \Cref{fig:figure1}C). 
More generally, there has been considerable interest in ciliary flows \cite{ding_mixing_2014,guzman-lastra_active_2021,ling_flow_2023} and the resulting particle capture \cite{ding_selective_2015}. 
Through a combination of flows, driven by the squid breath cycle and these ciliated surfaces, Vf aggregate on the Es light organ surface \cite{fung_vibrio_2024}.
The bacteria then swim along the surface of the light organ and through pores at the base of the appendages, enabled by chemotaxis.
This transit into the pores and into the light organ (marked P in \Cref{fig:figure1}C) marks a successful colonisation.

Over the last two decades, the bobtail squid has firmly established itself as a model system in symbiosis research \cite{nyholm_lasting_2021,visick_lasting_2021}.
This is due to the specificity of the relationship between Es and Vf, and the resulting cascade of physiological changes it precipitates in the squid \cite{peyer_eye-specification_2014,rader_persistent_2019}.
Es and Vf offer several additional desirable traits, making them well-suited to husbandry and culturing. 
The squid is relatively small and therefore suitable for lab husbandry, with juveniles measuring around 1.5 mm and adults growing to 35 mm. 
They mature rapidly, with a lifespan of around a year.
In addition to their light organ, female bobtail squid produce a supplementary reproductive organ, the accessory nidamental gland (ANG) \cite{mcanulty_failure_2023}.
The ANG stores a simple community of symbiotic bacteria that the squid requires for reproduction.
As such, one host offers researchers two regulated symbioses within a single model organism.
Despite the attention this system has received, many areas of study remain open for exploration.
One understudied area is the impact of changing climates, particularly thermal stresses, on these symbiotic relationships. 
Researchers have begun to study this relationship from the perspective of the symbiont \cite{cohen_adaptation_2019, cohen_adaptation_2020}, and more recently, the host \cite{Otjacques_Climate_2025}.
However, in general, it is not fully understood whether these symbiotic relationships pose hidden vulnerabilities or even confer resilience to changing climates. 
Several studies have linked external stresses to respiration in various cephalopods \cite{trubenbach_ventilation_2012}.
Prior work has shown that squid metabolic demands increase under stress, requiring ample oxygen to meet these additional demands \cite{demont_effects_1984}.
To compound the increased metabolic demands, thermal stress in cephalopods drastically impairs diffusion gradients required for oxygen uptake at the gills \cite{melzner_temperature-dependent_2006,zielinski_temperature_2001}. 
Organisms can combat these problems by increasing ventilation rates and internal fluid pressure. 
However, this increased respiration is metabolically taxing, driving a further need for oxygen, forming a negative feedback loop of great consequence for organisms living under changing climates \cite{rosa_synergistic_2008}.
The impact of these changes to ventilation and their downstream effects on symbiotic colonisation is currently an open question. 
In this study, we use mathematical modelling, computational simulations, and sensitivity analysis as a first step toward understanding this complex process. 

We develop a computational fluid dynamics model of early bacterial acquisition inside the squid using the Method of Regularised Stokeslets \cite{cortez_method_2001}.
Using this model, we assess, from the host perspective, the impact of temperature stress on early colonisation.
The specific way a squid breathes under thermal stress might inadvertently make it harder for its essential symbiotic partner to colonise it. 
Our model explores how fluid flow during respiration could drive these changes.
To understand how heat stress affects the establishment of symbiosis, we encode thermal stress effects into the model parameters to investigate their potential effects.
In particular, we examine the effects of breathing strength, breathing rate, ciliary flow strength, and internal cavity size using a sensitivity analysis framework.
By sampling this physiologically relevant parameter space, we examine the resulting variance in the time bacteria spend in regions crucial to the establishment of symbiosis.
We use a variance-based sensitivity analysis to quantify the individual and cooperative effects of these varying parameters on the temporal metrics of bacterial colonisation using Sobol indices \cite{sobol_global_2001}. 
Our results suggest that physiological alterations to breath cycles, particularly those anticipated due to thermal stress, can have a profound effect on the flow characteristics in the region of the light organ.
These resulting flows reduce the time bacteria spend in regions crucial to the early colonisation process, suggesting that thermal stress significantly impacts colonisation.

%% file: figures/figure1.tex
\begin{figure}[!t]
\centering
    \includegraphics[width=\linewidth]{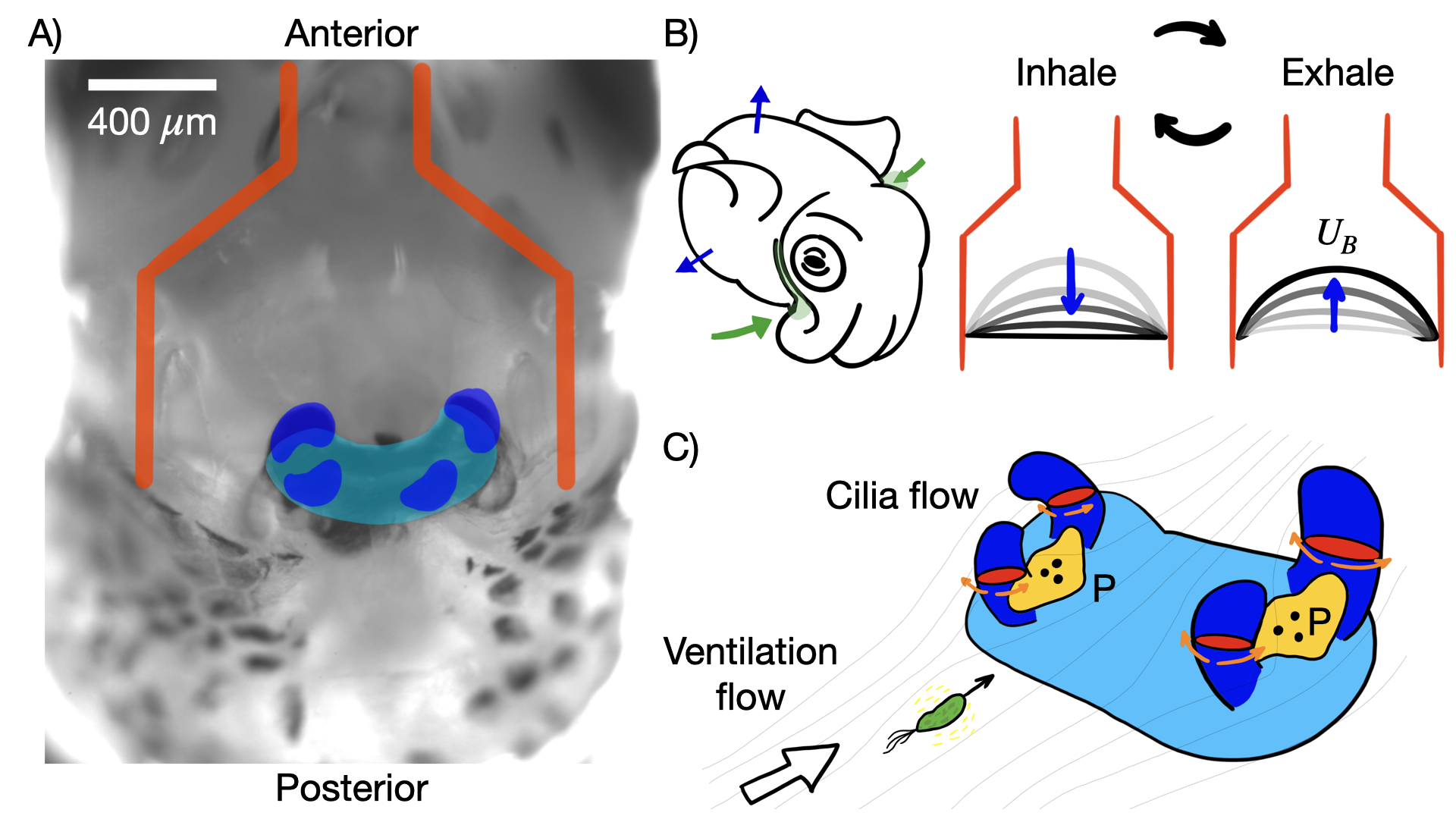}
    \caption{\textbf{Squid internal cavity.}
    A) A grey-scale experimental image of \textit{Euprymna scolopes} is shown. 
    Overlaid on the image, the external walls of the internal cavity, a funnel-like structure, are highlighted in red. 
    The body of the bi-lobed light organ is highlighted in light blue. 
    The individual appendages are highlighted in dark blue.
    B) A schematic of the internal flows and ventilation boundary conditions resulting from the expansion of the mantle within the squid during a breath cycle. 
    The expansion directions of the mantle during an inhale are shown in blue, and green arrows indicate where fluid enters the squid at the base of the mantle (left).
    A schematic of the Poiseuille flow boundary conditions at the inlet on the internal cavity, which models the breath cycle of the squid (right).
    C) The two major flow regimes experienced by bacteria within the cavity.
    The Poiseuille or ventilation flow dominates the far-field. 
    Close to the appendages, the beating of the cilia on their surface drives the flow.
    The bi-lobed light organ is highlighted in light blue, and 
    the cross sections (red) are those used in the two-dimensional simulations.
    Long cilia-covered regions are shown (dark blue), as well as short cilia regions (yellow), and the pair of three pores, through which the bacteria can pass into the internal ducts of the light organ, are marked (black, P).
    }
    \label{fig:figure1}
\end{figure}

%% file: methods.tex
\section{Model and Methods}

We first outline the mathematical model and computational methods used to simulate flows and bacterial motion in the squid.
Motivated by experimental measurements, we characterise the key features of the squid internal cavity. 
We then outline how to model the squid internal fluid flow using this characterisation.
Next, we describe how to simulate bacterial trajectories advected by these flows. 
We propose metrics based on the flow and bacterial trajectories to quantify the squid's ability to capture the bacteria. 

Our investigation identifies several key parameters of interest and physiological ranges in live squid that are likely to be strongly affected by thermal stress.
Finally, we discuss a method for conducting a sensitivity analysis using quasi-random sampling of the parameter space to disentangle the relative impacts of these key underlying parameters on our metrics. 

\subsection{Mathematical Model} \label{MathematicalModel}

\input{figures/figure2}

We first consider the fluid flow in a two-dimensional cross-section through the internal cavity of the squid, which intersects the light organ appendages, as shown in \Cref{fig:figure1} and \Cref{fig:figure2}. 
The fluid flow is driven by the expansion and contraction of the mantle as the squid breathes and cilia beat on the surface of the light organ; see \Cref{fig:figure1}B and \Cref{fig:figure1}C. 
The two-dimensional model captures the dominant flow dynamics (ventilation and ciliary) and allows for extensive parameter sweeps required for a robust sensitivity analysis, which is computationally prohibitive in three dimensions.

In prior work \cite{nawroth_motile_2017}, the flow around a pair of in vitro light organ appendages was studied experimentally and computationally. 
Our model generalises this previous work to include (1) the effects of the squid internal cavity, and (2) the temporal dynamics of respiration, and couples these with cilia-driven flows. 
This extension enables us to gain more realistic, detailed insights into the dynamics that occur \textit{in vivo}. 
This new model enables us to investigate the thermal stresses on the flow within the squid and their subsequent impact on the colonisation by Vf.
In particular, by exploring breathing regimes parameterised by the model, we conduct a sensitivity analysis to identify the breathing features that yield the most significant changes in bacterial trajectories.

The internal cavity has a few key features, as shown in \Cref{fig:figure1}.
We focus on modelling the funnel structure through which fluid passes as the squid breathes and the light organ (\Cref{fig:figure1}A). 
\Cref{fig:figure2}A shows the model system geometry used to approximate the internal structure of the squid in two dimensions.
\Cref{tab:Modelparameters} gives a list of the parameters used to characterise this system.

Experimental data have previously estimated the Reynolds number, the ratio of inertial to viscous stresses, of this system in the vicinity of the light organ to be on the order of $10^{-4}$ \cite{nawroth_motile_2017,gundlach_ciliated_2022}. 
As such, the velocity and pressure of the fluid flow within the system can be found by solving the Stokes equations, 
\begin{align}
- \nabla p(\bm{x},t)  + \mu \nabla^2 \bm{u}(\bm{x},t) &= \int_{\partial\Omega} \bm{f}(\bm{X}(\ell),t)\delta(\bm{X}(\ell) - \bm{x}) \, d\ell 
 \, , \label{StokesEquation_int}  \\
\nabla \cdot \bm{u}(\bm{x},t) &= 0 \, . \label{incompressibility}
\end{align}
In these equations, the fluid velocity, $\bm{u}(\bm{x},t)$, and pressure, $p(\bm{x},t)$ are a function of position, $\bm{x} = (x,y)$, and time, $t$, and the parameter $\mu$ is the fluid viscosity, assumed to be constant.
The temporally varying force density on the fluid from the squid, $\bm{f}(\bm{X}(\ell),t)$, is applied along the system boundaries, $\bm{X}(\ell)$, parametrised by $\ell$. 
Note, $\delta(\bm{X}(\ell) - \bm{x})$ is the two-dimensional Dirac delta function centred at location $\bm{X}(\ell)$. 

We prescribe boundary conditions to realise solutions for these equations within this system. 
The boundary conditions include a Poiseuille flow driven at the channel inlet, which models the squid respiratory ventilation (\Cref{fig:figure1}B). 
Furthermore, the surfaces of the appendages are set to have a non-zero tangential flow, which simulates the flow driven by ciliary beat waves (see \Cref{fig:figure2}B and \Cref{fig:figure2}C). 
The remaining internal cavity surfaces are subject to a no-slip boundary condition. 
Further details are given in Supplementary Material \Cref{SM:BOUNDARY}.

Zero net force and torque acting on the fluid from the squid, as a whole, is required, as in all biological systems in low Reynolds number environments. 
To enforce the net-zero constraints, we allow for background (far-field) translational and rotational velocities.

Given the fluid velocities on the boundaries and using the zero net-force and torque conditions, we can solve for the required force density on the boundaries, $\bm{f}(\bm{X}(\ell),t)$, by setting $\bm{x}=\bm{X}(\ell)$ in \Cref{StokesEquation_int} and \Cref{incompressibility}. 
We then use this force density in \Cref{StokesEquation_int} and \Cref{incompressibility}, to solve for the flow velocity, $\bm{u}(\bm{x},t)$, and pressure, $p(\bm{x},t)$, anywhere within the squid domain, $\bm{x}$. 

We then model the time-dependent dynamics of Vf bacteria by coupling them with the fluid flow within the squid.
We assume the bacteria are circular passive particles advected by the velocity calculated using the Stokes equations. 
We solve for the velocity at each position along the bacterial trajectory, then advance the bacteria position based on the resulting flow.
We include a (steric) contact force to account for the finite size of the bacteria interacting with the squid surfaces as they move through the system \cite{nawroth_motile_2017}. 
This force acts between the bacteria and the appendages within the light organ, causing local repulsion at close approach.
Hence, the advective motion of the bacteria is described by,

\begin{equation}
    \frac{d\bm{x}}{dt} = \bm{u}(\bm{x},t) + k \left( R + R_p - |\bm{x}_a - \bm{x}|\right) \frac{\bm{x_a} - \bm{x}}{|\bm{x_a} - \bm{x}|} H(R + R_p - |\bm{x_a} - \bm{x}|) \, , 
    \label{equationOfMotion}
\end{equation}
where $\bm{x}(t)$ is the location of the bacteria at time $t$, $\bm{u}(\bm{x},t)$ is the solution of the velocity from \Cref{StokesEquation_int} and \Cref{incompressibility}, the constant $k$ determines the scaling of the steric forces, $R$ and $R_p$ are the radii of the appendage and bacteria, respectively, and $\bm{x}_a$ is the location of the centre of the closest appendage to the bacteria. 
The Heaviside function, $H$, ensures that the perturbation to the motion due to the steric forces only acts on the bacteria when they are sufficiently close to the nearest appendage. 

\begin{table}[t!]
    \centering
    \begin{tabular}{|c|c|c|c|c|c|}
        \hline
        Symbol & Description & Value & Range & Units & Reference \\
        \hline
        \hline
               & Physical Parameters: Geometry and Flow & & & & \\ 
        \hline
        \hline
        $\ell_1$ & Wide section length & 867 & & $\mu$m & HHL \\
        \hline
        $\ell_2$ & Transition section length & 210 & & $\mu$m & HHL \\
        \hline
        $\ell_3$ & Narrow section length & 400 & & $\mu$m & HHL \\
        \hline
        $W_B$ & Bottom of channel width & 1000 & & $\mu$m & HHL \\
        \hline
        $W_T$ & Top of channel width & 400 & & $\mu$m & HHL \\
        \hline
        $\theta$ & Channel transition angle & 0.61 & & Rad & HHL \\
        \hline
        $\psi_0$ & Orientation of appendages & N/A & & Rad & \cite{nawroth_motile_2017} \\ 
        \hline
        $P_x$ & Appendage pair distance from centre line & 250 & & $\mu$m & HHL \\
        \hline
        $P_y$ & Appendage pair distance from channel entry & 495 & & $\mu$m & HHL \\
        \hline
        $\phi$ & Appendage pair angle to horizontal & 0.82 & & Rad & HHL \\
        \hline
        $R$ & Appendage cross-section radius & 45 & & $\mu$m & \cite{nawroth_motile_2017} \\
        \hline
        $R_{\text{eff}}$ & Appendage effective radius & 42.75 & & $\mu$m & \cite{nawroth_motile_2017} \\
        \hline
        $R_p$ & Bacteria radius & 1 & & $\mu$m & HHL \\
        \hline
        $d$ & Appendage separation & 41.75 & & $\mu$m & \cite{nawroth_motile_2017}  \\
        \hline
        $U_C$ & Maximum strength of cilia flow  & 600 & 400-800 & $\mu$m s$^{-1}$ & \cite{nawroth_motile_2017} \\ 
        \hline
        $U_B$ &  Maximum strength of Poiseuille flow & 125 & 50-200 & $\mu$m s$^{-1}$ & \cite{nawroth_motile_2017} \\
        \hline
        $\omega$ & Poiseuille flow frequency & 3 & 0.5-5.5 & $s^{-1}$ & \cite{visick_exclusive_2000}\\ 
        \hline
        $Y_0$ & Initial distance of bacteria from appendages & 225 & 135-315 & $\mu$m & \\
        \hline
        \hline
              & Computational Parameters & & & & \\ 
        \hline
        \hline
        $\epsilon$ & Regularisation parameter & 0.0167 & & $\mu\text{m}$ &  \\
        \hline
        $\rho$ & Discretization density & 30 & & $\mu\text{m}^{-1}$ & \\
        \hline
        $k$ & Scaling of steric forces & 20 &  & s$^{-1}$  &  \\
        \hline
    \end{tabular}
    \caption{\textbf{Model parameters}. 
    The first section presents the physical parameters related to the squid and the fluid flow, as shown in \Cref{fig:figure2}. 
    The second section presents the computational parameters necessary to conduct simulations. 
    Columns give the mathematical symbol used, a description of the parameter, the nominal experimental values in standard conditions, the range for parameters varied due to thermal stress, the units, and the reference for the nominal value. 
    HHL refers to values collected in the Heath-Heckman Laboratory for this study. 
    Values for $\psi_0$ are not provided here, as the parameter is set relatively in the code (see \cite{StephWilleniamsSquid}, based on \cite{nawroth_motile_2017}) 
    }
    \label{tab:Modelparameters}
\end{table}

\subsection{Numerical Methods} \label{NumericalMethods}

As outlined in the previous section, we solve for the forces exerted by the squid on the fluid, the background velocity and torque, and the velocity and pressure at any point within the cavity, given a specific set of boundary conditions on the squid model. 
We then use the velocity to solve for the bacterial trajectories within the squid.

We discretise the integral of the forces from the squid in the momentum equation, \Cref{StokesEquation_int}, as a sum of $N$ point forces,
\begin{equation}
\mu \nabla^2 \bm{u}(\bm{x},t) - \nabla p(\bm{x},t) + \sum_{i=1}^N \bm{f}_i(t)\delta(\bm{x}_i - \bm{x}) = \bm{0} \, . \label{StokesEquation}
\end{equation}
To solve for the velocity and pressure in our system, we apply the Method of Regularised Stokeslets to approximate the solution to the Stokes equations, \cite{cortez_method_2001,cortez_method_2005}. 
In the Method of Regularised Stokeslets, rather than working with point forces, we work with blobs, regularised forces.

In a reference frame of the squid mantle, the resulting relationship between the velocity and forces, as a solution to the Stokes equations, is given by,
\begin{equation}
    \bm{u}(\bm{x},t) = \sum_{i=1}^N \bm{S}^\epsilon(\bm{x},\bm{x}_i) \bm{f}_i(t)  - \bm{U}_0(t) - \bm{K}\Omega(t) \, . \label{flow_calculation}
\end{equation}
The right-hand side of the equation contains three distinct terms. 
The first of these contributions comes from $N$ regularised Stokeslets, $\bm{S}^\epsilon(\bm{x},\bm{x}_i)$, where the force blobs are distributed over the structures composing the mantle and the light organ appendages at points $\bm{x_i}$.
The discretisation spacing at which they are distributed along the boundaries is determined by the density of blobs per unit length, $\rho$, each having an associated size $\epsilon$.
The regularised Stokeslet, $\bm{S}^\epsilon(\bm{x},\bm{x}_i)$, allows us to calculate how each of these force blobs contributes to the flow at a location $\bm{x}$.
The term $\bm{U}_0(t)$ sets the translational flow in the far-field of the system and the vector $\bm{K} = (-y,x)^T$ encodes the non-trivial contributions of the torque $\bm{\tau}(t) = \bm{x}\times\bm{\Omega}(t)$, where $\bm{\Omega} = (0,0,\Omega(t))$. 
These two terms are included to enforce a net-zero force and net-zero torque constraint on the system.
These terms describe the squid translational and rotational motion in the lab frame, respectively.
If we know velocity boundary conditions at sufficiently many points at a given time, as in our system, we can calculate the forces required on the boundaries to enforce them.
Then, using the forces, the velocity is solved for at any point within the domain, $\bm{x}$, using the same equation. 
See Supplementary Material \Cref{SM-MRS} for further details on applying the Method of Regularised Stokeslets to solve for the fluid flow within the squid from the temporally varying boundary conditions. 

Finally, to determine the bacterial trajectories, we numerically integrate \Cref{equationOfMotion}. 
We use the Dormand-Prince Runge-Kutta method, implemented in MATLAB \cite{shampine_matlab_1997}, to perform this calculation.
Since the boundary conditions are time-dependent, we solve for the forces and flows using \Cref{flow_calculation} at each time step, which are then used in \Cref{equationOfMotion}.

\subsection{Parameters and Metrics} \label{Metrics}

To study the impacts of thermal stress on the entire system, we select parameters most likely to be affected by thermal stress.
We simulate the trajectories of bacteria as these parameters are varied.
The first three parameters we choose to vary are the ventilation strength or the maximum strength of the Poiseuille flow, $U_B$, the breathing frequency or the Poiseuille flow frequency, $\omega$, and the maximum strength of cilia flow, $U_C$. 
The parameter values and ranges are given in \Cref{tab:Modelparameters}.
Additionally, we vary the initial distance between the bacteria and the appendages, $Y_0$. 
We seed 100 bacteria uniformly in the $x$-direction at a distance $Y_0$ in the $y$-direction from the appendages (as shown in \Cref{fig:figure2}A).
Only half, 50, of these bacterial trajectories are simulated, since the flow dynamics are symmetric around $x = 0$.
Varying $Y_0$ allows us to sample the effects of bacteria arriving at our initial condition at different phases of the breath cycle and to capture variations in the length of the internal cavity of the squid.
Supplementary Material \Cref{SM-PARAMS} provides a discussion of the chosen ranges for the varied parameters.

To analyse the set of simulated bacterial trajectories, we use metrics to quantify each trajectory with a single scalar value.
In our analysis, we focus on two such metrics: the average total time per bacteria spent close to an appendage, in a critical zone, over ten breath cycles, $\tau_{tot}$, and the longest continuous time a single bacterium is close to an appendage, in this critical zone, over ten breath cycles, $\tau_{max}$.
Close bacteria are those within $67.5~\mu$m of one of the four appendages in the system.
Vf bacteria swim at a speed of approximately $45~\mu\text{m s}^{-1}$ \cite{Zhuang2024}. 
Thus, this distance would take at least 1.5 seconds for Vf to traverse.

As we will see in \Cref{sec:extremes}, the breath cycle can recirculate bacteria under certain circumstances. 
Hence, the total time, $\tau_{tot}$, gives a measure of time spent in the critical zone, accounting for possible periodic returns.
If this total time is sufficiently long, the fluid flow during respiration will give bacteria a good chance to reach their target destination and colonise the squid by being (possibly repeatedly) deposited close to the appendage, regardless of the role of bacterial swimming.
The most prolonged period a bacterium spends close to an appendage in the critical zone, $\tau_{max}$, allows us to assess the potential effects of the bacterium's autonomy.
For a bacterium spending sufficiently long close to an appendage, we expect that it has a good chance of reaching its target destination and colonising the squid through its motility. 

\subsection{Sensitivity Analysis} \label{SensitivityAnalysis}

To understand how different parameter values that encode breathing dynamics affect colonisation, we perform a sensitivity analysis.
In this analysis, we generate an extensive collection of parameter sets, each of which is independently drawn from a range about the nominal values, see \Cref{tab:Modelparameters}.
Using these parameter sets, we then simulate fluid flows and bacterial trajectories driven by the associated breathing dynamics and measure the resulting colonisation metrics.
By assessing the relationships between the parameters and the calculated metrics, we understand how changes in breathing dynamics affect colonisation.
In particular, we quantify the variance of the metrics, $\tau_{tot}$ and $\tau_{max}$, across the parameter space.
To ensure good coverage of the phase-space of our parameters, we use Saltelli's extended Sobol sample algorithm to generate 81,920 parameter sets \cite{sobol_global_2001,saltelli_making_2002,saltelli_variance_2010}, which exceeds the suggested sample size for four parameters \cite{zhao_modeling_2023}. 
Simulations of these parameter sets enable us to capture the highly nonlinear dependence of the output metrics on the parameter inputs.
We used the open-source Python library SALib \cite{herman_salib_2017} to generate the sample parameters and perform the subsequent Sobol analysis. 
Sobol analysis allows us to understand which parameters are responsible for the variance in the metrics $\tau_{tot}$ and $\tau_{max}$.
The analysis decomposes the total variance of each metric into first-order indices, arising solely from changes in a single parameter, and second-order indices arising from pairwise parameter interactions.
We also present the total order index for each parameter, how much of the variance in a metric is due to that parameter, including all interactions with other inputs.
All code and data for the research conducted here can be found in a corresponding GitHub repository \cite{StephWilleniamsSquid}.

%% file: figures/figure2.tex
\begin{figure}[!t]
\centering
    \includegraphics[width=\linewidth]{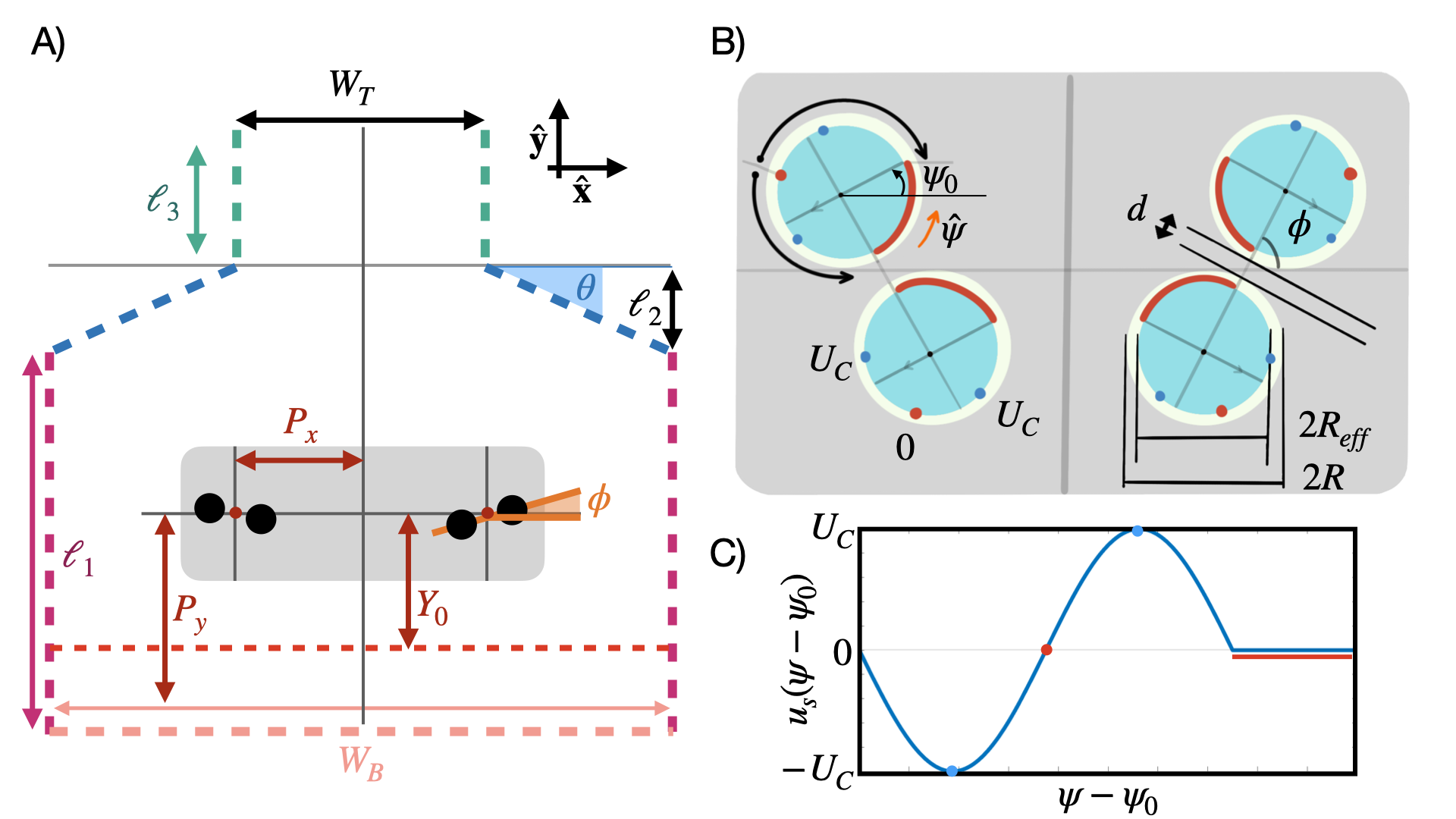}
    \caption{\textbf{Model schematic.} 
    A) Schematic of the model of the internal cavity, a two-dimensional channel geometry intersecting through the squid, see Figure \ref{fig:figure1}. 
    Slice through the light organ is shown in grey, and cross sections of the appendages are shown in black. 
    The initial distance of Vf from the appendages is shown as a red dashed line. 
    Relevant measurements required to characterise this internal channel of the squid are labelled. 
    B) A close-up of the appendage pairs is shown.
    Top left: The flow directions (black arrows), driven by the beating of the cilia, on the appendage surface are shown, as well as the angle $\psi_0$ which defines the orientation of each appendage in the $\hat{\psi}$ direction. 
    Bottom left: Key flow speeds of average beating cilia are shown. 
    Top right: Appendage spacing, $d$, and the relative angle, $\phi$, between the appendage pairs are defined.
    Bottom right: The radius, $R$, at which steric (contact) interactions occur is defined, and the effective radius, $R_\text{eff}$, at which the Stokeslets are placed. 
    C) Magnitude of the tangential flow along the surface of one of the appendages. 
    Blue and red dots along this curve correspond to those in Figure 2B.
    }
    \label{fig:figure2}
\end{figure}

%% file: results.tex
\section{Results}

\subsection{Model Dynamics}

\input{figures/figure3}

We first present the model dynamics, with the varying parameters set to their nominal experimental values, as given in \Cref{tab:Modelparameters}.

Simulations of the time-dependent flow field in the squid frame of reference are displayed in \Cref{fig:figure3}.
We present snapshots of the flow field at various time points along a breath cycle (\Cref{fig:figure3}A).
Throughout the breath cycle, the flow is always fastest at the appendage surfaces.
During the breath cycle, the fastest flow occurs at the apex of an exhalation (\Cref{fig:figure3}B), where flows at the outlet (top) reach approximately $300~\mu\text{m}\text{ s}^{-1}$.
Furthermore, at this point in the breath cycle, upstream of the outlet (above the appendages) adopts a profile that is nearly Poiseuille, and the fluid generally moves toward this outlet.
Flow redirection occurs near the appendages, where two vortical regions are present upstream.
Despite this local redirection, the flow is directed predominantly to the outlet.
As we progress in time through the breath cycle toward the inhalation phase (\Cref{fig:figure3}C-E), the flow near the outlet gets weaker, and two vortices appear downstream of the appendages.
At the point of inhalation (\Cref{fig:figure3}F), the ventilation flow is nearly absent, and fluid is no longer directed towards the outlet. 
Instead, at this time, the flow direction is briefly reversed in regions away from the vortices.
From this first simulation, we observe that it is necessary to model the temporal dynamics of the fluid flow within the internal cavity when considering bacterial trajectories and the resulting colonisation, as flows vary significantly throughout the breath cycle. 

\input{figures/figure4}

Next, we present the trajectories of bacteria in the absence and in the presence of breathing dynamics.
In \Cref{fig:figure4}A, time-independent boundary conditions are applied, using a fixed magnitude Poiseuille flow as the boundary condition on the inlet, to solve for the fluid flow. 
This flow profile corresponds to that present in \Cref{fig:figure3}B, modelling a constantly exhaling squid. 
We then use this fluid flow to solve for the bacterial trajectories using \Cref{equationOfMotion}.
We observe that the bacteria initially move through the system along relatively straight trajectories.
Once they reach the region where the appendage boundary conditions dominate the flow, the bacteria are redirected.
They circulate around vortices in the flow (VZ1 in \Cref{fig:figure4}A), before moving through a stagnant zone.
The stagnant zone (SZ in \Cref{fig:figure4}A) was previously reported by \cite{nawroth_motile_2017} and is present between each of the appendage pairs. 
In this stagnant region, flow speeds are drastically slower than in the surrounding areas. 
Once bacteria reach the stagnant zones, where they are no longer strongly advected by the ventilation flows, they can perform chemotaxis and enter the pores (P in \Cref{fig:figure1}) into the light organ, successfully colonizing the squid. 
Two other vortices, VZ2, are also observed downstream of the appendages, and the bacteria do not enter these regions when considering a time-independent flow within the squid. 

In contrast, \Cref{fig:figure4}B shows the trajectories of a collection of bacteria advected by the time-dependent boundary condition that models the squid breathing, \Cref{timeDependenceVent}. 
This time dependence manifests as ripples in the trajectories, particularly upstream of the appendages.
At inhalation, due to the reversal of flow (\Cref{fig:figure3}F), bacteria away from the appendages have their direction reversed and often end up circling within the cavity.
Specifically, the flows drive more mixing upstream, with bacteria circling repeatedly in the stagnant zones and vortical zones near the appendages.
Further, when the bacteria enter VZ2, they are reversed, with bacteria sometimes passing close to the appendages again.
As a result, the bacteria are redeposited multiple times in the appendage area, close to the pores, before exiting.
From this initial comparison of trajectories, we observe that understanding the temporal dynamics is essential to determining the time Vf spends near the Es appendages and pores, and, therefore, the resulting colonisation process.

\input{figures/figure5}

In \Cref{Metrics}, we proposed two metrics to quantify the ability of the squid to be colonised by the bacteria: the first measures the average time spent per bacterium at a critical distance from the appendages, $\tau_{tot}$, and the second measures the longest continuous time spent in the critical zone, $\tau_{max}$. 
We choose a threshold distance and the corresponding critical zone to include the stagnant zones (\Cref{fig:figure4}) and the pores (\Cref{fig:figure1}), from where bacteria can easily swim and enter and colonise the squid light organ. 
\Cref{fig:figure5} presents this quantification for the trajectories shown in \Cref{fig:figure4}B, using nominal experimental values for the model parameters. 

\Cref{fig:figure5}A presents the trajectories of bacteria, initially seeded uniformly at the inlet, overlaid with the critical zones. 
Darker shades of blue indicate areas within Es where the Vf occupy for more extended periods.
The yellow annuli, centred on the four appendages, show the threshold distance and corresponding critical zone used for the metrics.
\Cref{fig:figure5}B presents the times at which each of the bacteria is within the critical zone. 
Bacteria that start close to the centre line of the system, near $X_0=0$ $\mu$m, quickly reach the critical zone.
Many of these bacteria, particularly those starting close to $X_0=\pm270$ $\mu$m, re-enter this zone repeatedly at later times. 
The bacteria initially located near the exterior walls of the cavity ($X_0=\pm540$ $\mu$m) typically take longer to reach the critical zone, but eventually do so.

\Cref{fig:figure5}C and \Cref{fig:figure5}D present the two metrics for each bacterium in this nominal simulation.
\Cref{fig:figure5}C presents the total time each bacterium spends in the critical zone near the appendages.
Averaging all these values yields the average time per bacterium spent in this critical zone, $\tau_{tot} = 0.390$ s.
\Cref{fig:figure5}D presents the longest consecutive time each bacterium spends in a critical zone.
We measure the maximum time among all bacteria to give our second metric, $\tau_{max} = 0.491$ s.
Using these metrics to analyse realistic simulations that account for the internal cavity and breathing dynamics, we assess the impact of varying the parameters affected by thermal stress in the next section. 

\subsection{Effect of Varying Parameters on Bacterial Trajectories and Colonisation Metrics} \label{sec:extremes}

In this section, we present the effects of varying the four parameters expected to be affected by thermal stress in the squid (see \Cref{Metrics}) on bacterial trajectories and the corresponding metrics. 

\input{figures/figure6}

We begin by presenting the bacterial trajectories resulting from univariate extremes of the parameter range, with all other parameters held at their nominal values (\Cref{fig:figure6}A-D). 
In \Cref{fig:figure6}A, we observe the effect of the Poiseuille ventilation flow strength, $U_B$, varying between strong and weak breathing dynamics. 
When ventilation flow is weak, alternating between two vortices dominate the bacteria paths over time, and the bacteria do not explore much of the cavity.
When ventilation flow is strong, trajectories exhibit a more chaotic mixing pattern, with bacteria travelling closer to the appendages and a larger proportion passing through the stagnant zone.
\Cref{fig:figure6}B illustrates the impact of breathing frequency, $\omega$, where a lower breathing rate yields smoother trajectories closer to the appendages, and the bacteria travel through more of the internal cavity. 
A higher rate results in less smooth trajectories and fewer bacteria travelling through the critical zone near the appendages.
\Cref{fig:figure6}C shows the effect of the maximum cilia-driven flow strength on the appendages, $U_C$.
Stronger ciliary flows overpower the ventilation flow, leading to greater backflow and more bacterial circling.
Finally, in \Cref{fig:figure6}D, we observe the effects of the initial position of the bacteria, $Y_0$.
This parameter has a less visible impact on the trajectories than the other three. 
We observe that bacteria initialized closer to the appendages explore a larger area in the vortical zones upstream (VZ2).
This slight difference indicates that the ciliary flows more readily recirculate these closer bacteria.

To better understand colonisation within the Es-Vf system and to quantify trends in the data as we vary the four parameters over their entire ranges, we present heatmaps of the metrics $\tau_{tot}$ and $\tau_{max}$ versus pairwise parameters. 
Select pairwise parameter heatmaps are shown in \Cref{fig:figure6}E-G.
The blue regions correspond to parameters and corresponding bacterial trajectories that yield metric values lower than those obtained with the nominal parameter values, suggesting that colonisation will be more challenging in these parameter regimes. 
Conversely, the yellow regions correspond to parameters and trajectories with times exceeding the metrics for the nominal values, suggesting that colonisation will occur more readily in these cases. 
All pairwise parameter heatmaps for the four varying parameters, with further details describing the analyses to create the heatmaps, are presented in Supplementary Material \Cref{SM-SCATTER}.

We first present results for varying breathing strength, $U_B$, and frequency, $\omega$, as these two parameters are highly affected by thermal stress, as discussed further here and in \Cref{sec:importance}. 
\Cref{fig:figure6}E presents the total time spent in the critical zone (near the appendages, as shown in \Cref{fig:figure5}A), and \Cref{fig:figure6}F presents the maximum continuous time spent in the critical zone. 
We observe that the total time (\Cref{fig:figure6}E) is generally equal to or greater than the nominal total time (i.e., when the squid is not stressed), except when both strength and frequency are large. 
For both of these parameters to be significantly larger than nominal values would be anomalous, as the fluid flux through a breath cycle is limited for an individual squid, see Supplementary Material \Cref{SM-PARAMS}. 
In comparison, for shallow breathing, when the ventilation strength is low and the frequency is high, we observe in \Cref{fig:figure6}F that the bacteria spend significantly less continuous time, $\tau_{max}$, in the critical zone.
Thermal stresses, when we expect breathing changes such as shallow breathing, have been shown to disrupt the establishment of symbiosis \cite{Otjacques_Climate_2025}.
This result for shallow breathing suggests a physical mechanism that reduces colonisation by reducing the maximum continuous time available to bacteria to enter the squid.
Furthermore, we note that slower breathing (smaller $\omega$) allows bacteria to spend a more extended period close to the appendages, regardless of ventilation strength ($U_B$), as observed in the trajectories at the extremes of breathing frequency (\Cref{fig:figure6}B). 
This is in contrast to what is observed for strong breathing (large $U_B$) in other cases (see \Cref{fig:figure6}G and Supplementary Material \Cref{SM-SCATTER}), when strong breathers achieve lower total times in the critical zone, suggesting that slower breathing compensates for stronger breathing. 

In \Cref{fig:figure6}G, we present the average total time in the critical zone, $\tau_{tot}$, for varying breathing strength, $U_B$, and initial bacterial distance from the appendages, $Y_0$. 
We observe nonlinear behaviours for both $\tau_{tot}$ and $\tau_{max}$ (see Supplementary Material \Cref{SM-SCATTER}). 
In general, weaker breathing (lower $U_B$) in longer squid cavities (larger $Y_0$) results in shorter times for both metrics, disrupting colonisation. 
The strongest breathers, regardless of the size of the internal cavity, generally achieve lower total times in the critical zone, $\tau_{tot}$, while having nominal values of $\tau_{max}$. 
Note that this is in contrast to what may be expected from the observations of the maximum extreme value trajectories in \Cref{fig:figure6}A. 
When the breathing strength is at its nominal value, $U_B = 125\mu\text{m s}^{-1}$, the value observed when the squid is not stressed, we observe long total and continuous times in the critical zone, regardless of the cavity length. 
This suggests that the nominal breathing strength is ideal for colonisation, since squid sizes vary. 

The plots where the strength of the ciliary flows ($U_C$) is varied are generally linear in the other parameter (see Supplementary Materials \Cref{SM-SCATTER}). 
These results show that the ciliary flows are generally unable to overcome differences in breathing strength, frequency, and the initial bacterial distance from the appendages, independently.

These data highlight how stress on the squid, leading to changes in ventilation and resulting fluid flows, can affect colonisation processes. 
The relationships between the temporal metrics measuring colonisation and the varying parameters are complex and highly nonlinear. 
In the next section, we will explore this further and present additional results on how thermal stress will negatively impact Es colonisation by Vf. 

\subsection{Importance of Varying Parameters on Colonisation Metrics} 
\label{sec:importance}

\input{figures/figure7}

Conducting simulations and analysing our data as presented in \Cref{sec:extremes}, we examine how parameter variations affect the resulting bacterial trajectories within the squid and the corresponding metrics for bacterial times in the critical zone, thereby improving our understanding of the colonisation process of the Es-Vf system.
In this section, we quantify the relative importance of each parameter on the metrics via sensitivity analysis and present the data, considering the impacts of thermal stress on the squid. 

The probability distribution functions (PDF) of $\tau_{tot}$ and $\tau_{max}$ are shown in \Cref{fig:figure7}A and \Cref{fig:figure7}B, respectively, for three different sets of data. 
The baseline dataset includes all simulations in which all four parameters, $U_B$, $\omega$, $U_C$, and $Y_0$, are varied independently, as given in \Cref{tab:Modelparameters}.
We explore trends in this data with consideration of thermal stress, considering two extreme breathing regimes: a deep-breathing dataset, which includes all data with high strength and low frequency, and a shallow-breathing dataset, which consists of all data with low strength and high frequency.
We already noted above in the discussion of Figure \ref{fig:figure6}F that shallow breathing, expected to be observed when squid are under stress, tends to result in less continuous time in the critical zones, negatively impacting colonisation, whereas slower breathing results in increased continuous time, facilitating colonisation. 
We now further quantify the understanding of the metrics and the colonisation process using these PDFs. 

In \Cref{fig:figure7}A, we present the PDFs for the average total time per bacterium spent in the critical zone near the appendages, $\tau_{tot}$.
We observe a peak in the baseline dataset at approximately $0.32$ s, which is below the baseline mean of $0.38$ s.
The deep breathing data has a slightly higher mean and a peak at about $0.4$ s, followed by a rapid drop-off.
In contrast, the shallow breathing data has a weaker peak at 0.27 s, but due to its heavier right-tail, it has a mean of $0.4$ s.
This right tail is likely due to the flow recirculation dynamics, which is more common in flow regimes with low $U_B$ (see \Cref{fig:figure6}A).
This suggests that when shallow breathing occurs, more bacteria spend less total time in the critical zone, and the flow recirculation leads to greater data variance and so less consistent colonisation. 
In comparison, deep breathing leads to more consistent bacterial trajectories with more total time spent in the critical zone.
Note that the conclusions here are different than the results presented in \Cref{fig:figure6}. 
The dataset and analysis here are more inclusive, as they include all simulations, whereas in \Cref{fig:figure6} only individual parameters or parameter pairs are varied, with all others held at their nominal values. 

In \Cref{fig:figure7}B, we plot the PDFs for the continuous maximal time a bacteria spends in the critical zone. 
This metric is significant for the colonisation process, as it measures the time the bacteria have in the critical zone to swim to the pores using chemotaxis and into the squid light organ. 
For the baseline data, we observe a clear peak in $\tau_{max}$ at approximately $0.5$ s. 
However, in addition to this peak, a lower, less prominent peak is present at approximately $0.35$ s.
The PDF of our shallow-breathing metric, $\tau_{max}$, shows a strong peak at $0.35$ s and no clear peak at the mean of $0.4$ s. 
In contrast, the deep breathing data has a much longer continuous maximum time, $\tau_{max}$, with the peak of the pdf at $0.54$ and a mean of $0.59$ s.
These data further support the above explanation that physical processes reduce the continuous time available for Vf to swim into the light organ when the squid are thermally stressed, thereby disrupting the establishment of symbiosis.

Due to the nonlinear dependence of the metrics on the parameters, Sobol sensitivity analysis is an appropriate tool for decomposing the impacts of parameter variation on the metrics. 
\Cref{fig:figure7}C and \Cref{fig:figure7}D show the first (blue) and total (orange) Sobol indices from this analysis, for $\tau_{tot}$ and $\tau_{max}$, respectively, using the baseline data, consisting of all simulations of the four varying parameters. 
The second-order Sobol indices from this analysis are presented in Supplementary Material \Cref{SM-SOBOL2}.

\Cref{fig:figure7}C presents how much of the variance in $\tau_{tot}$ is due to the varying parameters, see \Cref{SensitivityAnalysis} for more details. 
We observe that the total variation (orange) of the breathing strength, $U_B$, the breathing frequency, $\omega$, and the initial distance the bacteria are from the appendages, $Y_0$, significantly contribute to the variance in $\tau_{tot}$.
This suggests that changes in these three parameters due to stress on the squid will result in large changes in bacterial trajectories and in the total time the bacteria spend in the critical zone. 
These changes in the metric will be due to changes in that single parameter and to its higher-order interactions with other parameters. 
These effects are particularly pronounced for $U_B$ and $Y_0$, which show the largest total and first-order indices. 
This result indicates that independent changes in the breathing strength, $U_B$, and the squid size, $Y_0$, will significantly alter the total time spent in the critical zone. 
In \Cref{fig:figure6} and Supplementary Material \Cref{SM-SCATTER}, we have presented in more detail how $\tau_{tot}$ depends on these parameters, see the discussion in \Cref{sec:extremes}.
The values of the second-order indices, as presented in Supplementary Material \Cref{SM-SOBOL2}, show that only interactions between $U_B$ and $Y_0$ contribute significantly to the variance observed. 

For $\tau_{max}$ presented in \Cref{fig:figure7}B, all of the parameters collectively contribute significantly to the variance in the metric, since we have significant total indices for each parameter.
However, the only parameter with significant first-order contributions to the variance in $\tau_{max}$ is $\omega$, the breathing rate. 
We observed in \Cref{sec:extremes} the importance of breathing frequency, specifically that slow breathing can overcome strong breathing in this colonisation metric (\Cref{fig:figure6}F).
No second-order interactions are of explicit significance (Supplementary Material \Cref{SM-SOBOL2}), meaning the remainder of the variance must come from the collective small contributions of higher-order effects.

From these results, we observe that both metrics are somewhat robust to univariate parameter changes.
However, thermally stressed squid are expected to exhibit breathing patterns with multiple feature changes as shallow breathing involves both a reduction in breath strength and an increase in breath cycle frequency \cite{demont_effects_1984}.
These results suggest that changes across multiple aspects of breathing dynamics can drive significant alterations in the time bacteria spend in areas critical to symbiosis formation, as would occur under thermal stress.

%% file: figures/figure3.tex
\begin{figure}[!t]
\centering
    \includegraphics[width=\linewidth]{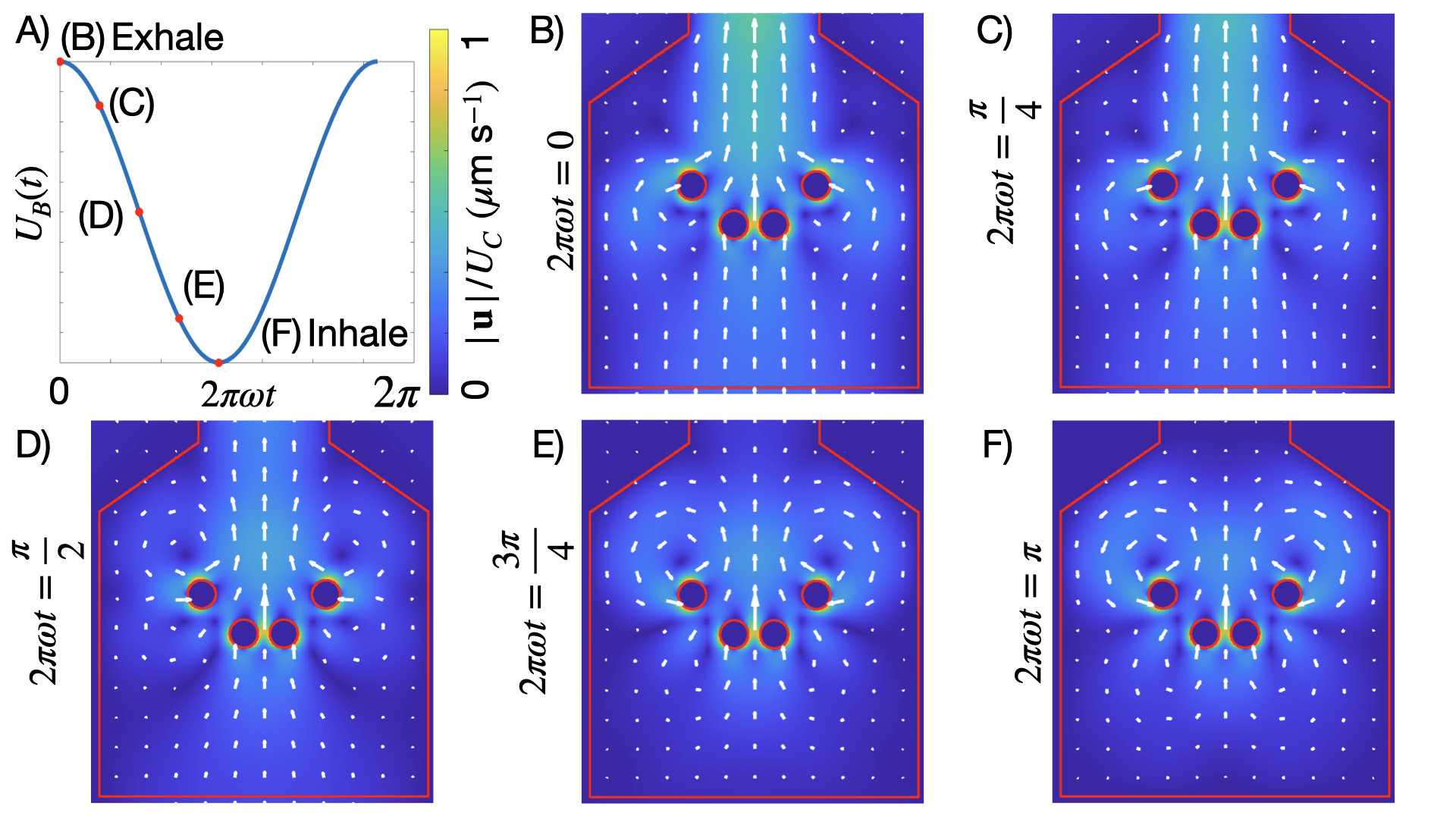}
    \caption{\textbf{Flow fields during a breath cycle in the internal cavity.} 
    A) The maximum magnitude of the temporally-varying Poiseuille flow boundary condition, \Cref{timeDependenceVent}.
    Five points are indicated along half of the breath cycle,
    from an exhale (where the ventilation flow at the inlet is maximal) to an inhale (where the ventilation flow at the inlet is zero). 
    The dynamics are symmetric in time about the midpoint of the breath cycle.
    B-F) The panels present the flow field at the highlighted time points in (A). 
    The squid internal cavity boundaries are shown in red.
    The local colour of the background presents the magnitude of the normalised local flow, scaled to the colour bar given in (A). 
    }
    \label{fig:figure3}
\end{figure}

%% file: figures/figure4.tex
\begin{figure}[!t]
\centering
    \includegraphics[width=\linewidth]{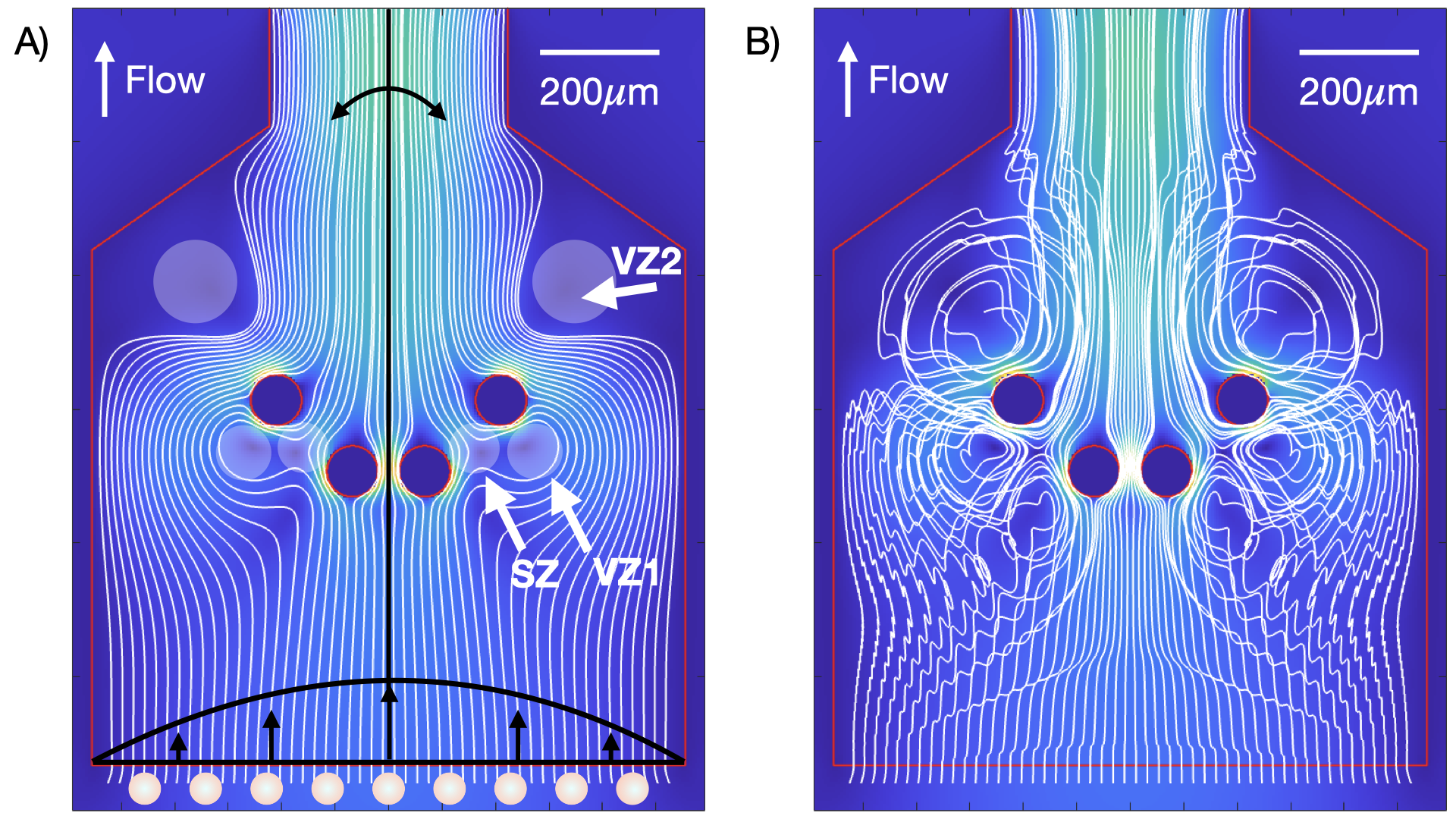}
    \caption{\textbf{Bacterial Trajectories.} 
    A) The trajectories for a set of bacteria in a time-independent flow field. 
    Schematics of bacteria are shown at the initial position (not to scale). 
    One hundred bacteria are initially seeded uniformly on this line.
    The flow is time-independent, with the ventilation boundary condition set using \Cref{timeDependenceVent} with $2\pi\omega t = 0$ and a maximum strength of $U_B = 125\mu\text{m s}^{-1}$ (shown at the bottom in black) remaining constant.
    Several regions of interest are highlighted:
    Stagnant zones (SZ), where the fluid velocity is low, vortical zones close to the appendage pairs (VZ1), and vortical zones downstream of the appendages (VZ2).
    B) The trajectories for a set of bacteria in a time-dependent ventilation flow. 
    The same number of bacteria are initialised in the same location as in (A). 
    The time-dependent ventilation boundary condition at the inlet is set using \Cref{timeDependenceVent}, with a maximum strength of $U_B = 125~\mu\text{m s}^{-1}$, and the resulting fluid flow is presented in \Cref{fig:figure3}.
    The trajectories span 14 breath cycles. 
    In (A) and (B), the background colour indicates the local flow velocity (yellow for the fastest flows and blue for the slowest).
    }
    \label{fig:figure4}
\end{figure}

%% file: figures/figure5.tex
\begin{figure}[!t]
\centering
    \includegraphics[width=\linewidth]{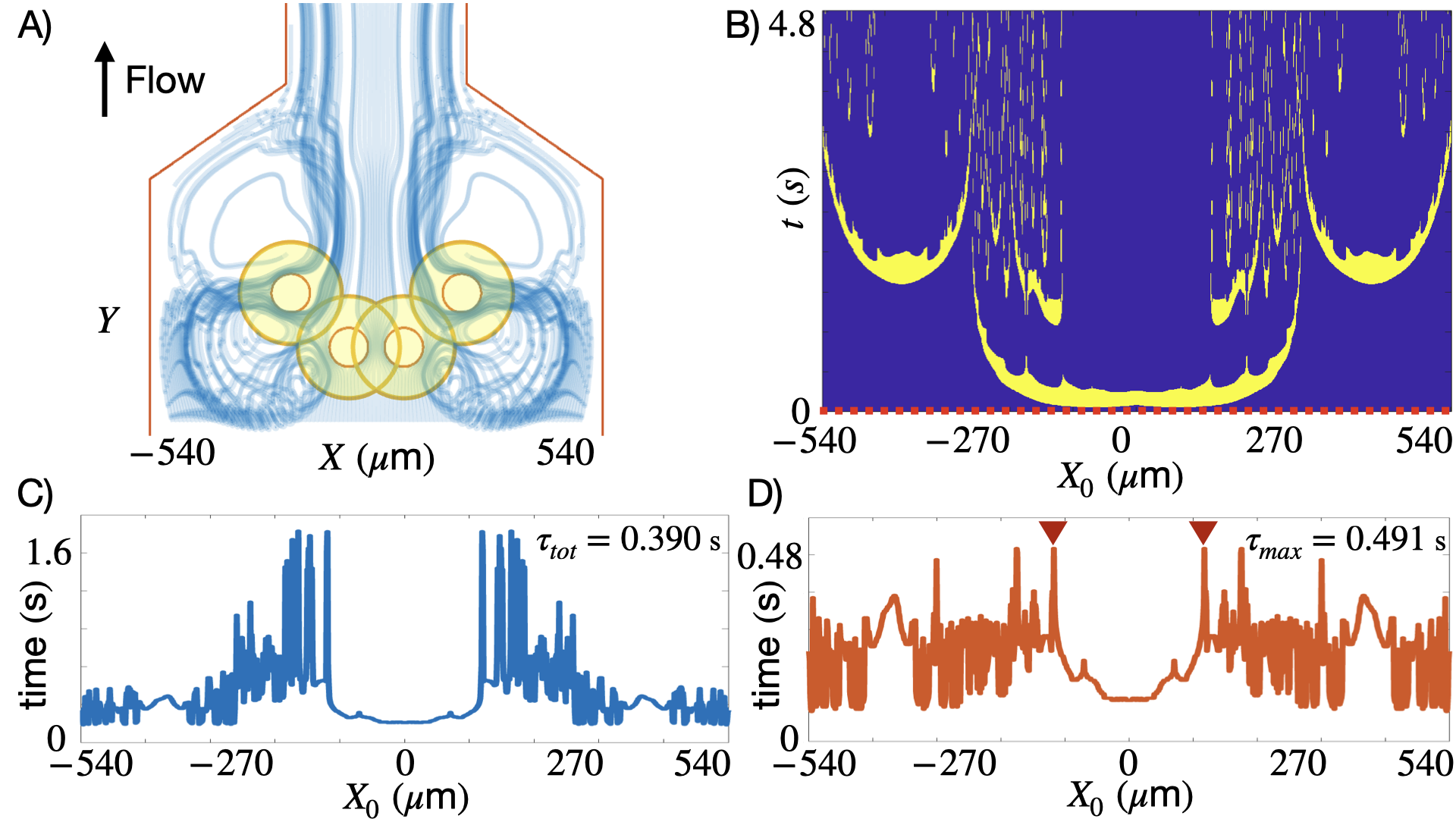}
    \caption{\textbf{Critical zone metrics.} 
    A) Collection of bacterial trajectories, as in \Cref{fig:figure4}B, overlaid by the critical zone used for the metrics, $\tau_{tot}$ and $\tau_{max}$. 
    One hundred bacteria are initially seeded at the bottom of the channel. 
    The system boundaries are shown in red, and the trajectories are shown in blue for 14 breath cycles.
    The four yellow annuli show the critical distance and corresponding zone used for metric calculation.
    B) Heatmap of bacteria proximity to the appendages as a function of time for given initial positions of bacteria, $X_0$, on the dotted red line.
    When a coordinate is yellow, it indicates that the bacteria are in the critical zone (yellow regions in (A)). 
    C) The total time each bacterium, initialised at $X_0$, is in the critical zone (yellow regions in (A)).
    The mean of these values, $\tau_{tot}$, is given in the inset.
    D) The maximum continuous time each bacterium, initialised at $X_0$, spends in the critical zone (yellow regions in (A)).
    The maximum across all the bacteria is marked with dark red arrows, and the corresponding value, $\tau_{max}$, is given in the inset.
    }
    \label{fig:figure5}
\end{figure}

%% file: figures/figure6.tex
\begin{figure}[!t]
\centering
    \includegraphics[width=\linewidth]{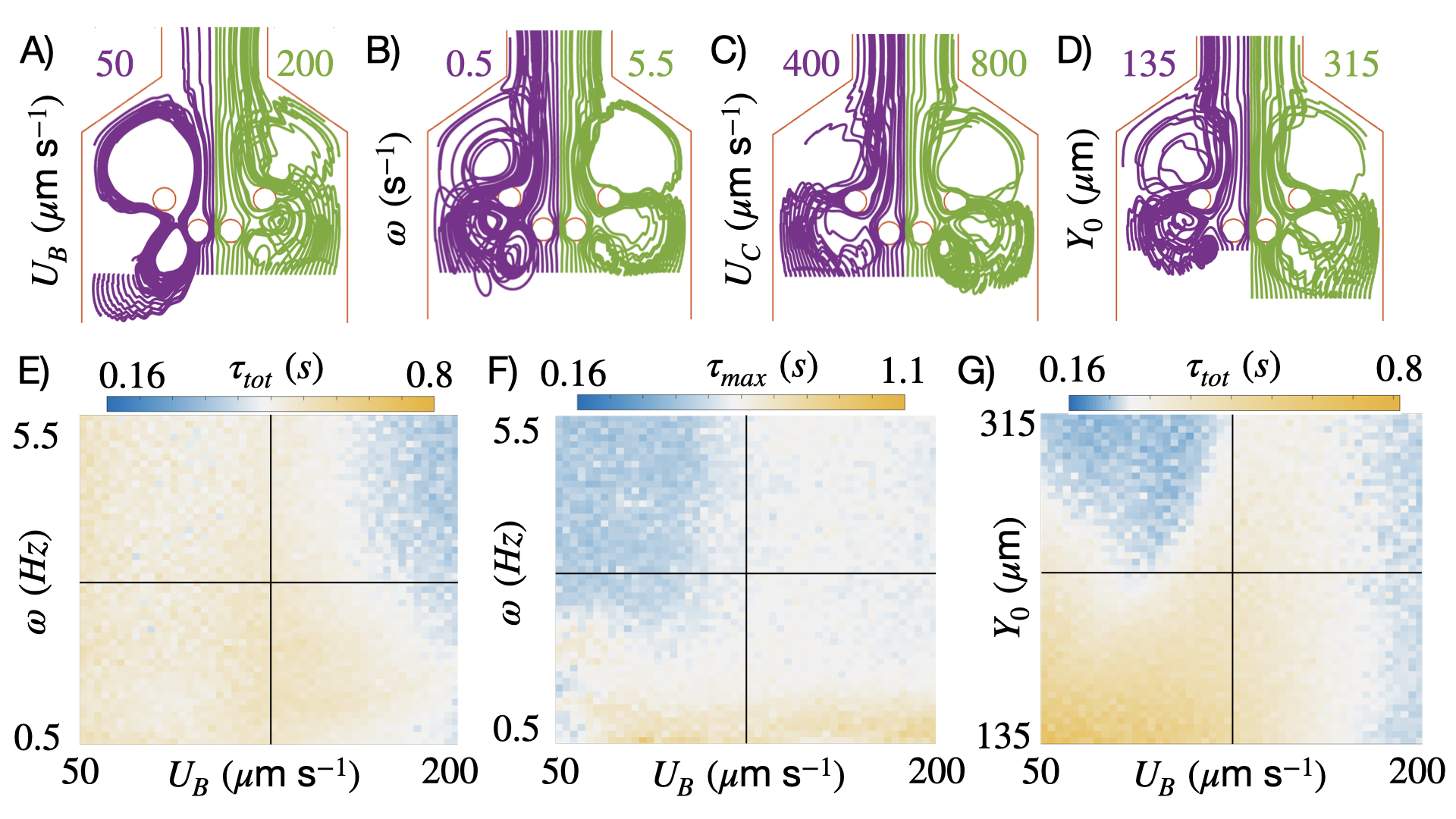}
    \caption{\textbf{Varying Parameters.} 
    A-D) 
    Bacterial trajectories resulting from univariate extremes of the four parameters: (A) the ventilation strength, $U_B$, (B) the breathing frequency, $\omega$, (C) the maximum strength of cilia flow on the appendages, $U_C$, and (D) the initial distance the bacteria are from the appendages, $Y_0$.
    Purple trajectories (left) are the simulations in which the modified parameter is set to its minimum, and green trajectories (right) present the trajectories from the maximum value. 
    All other parameters are set to their nominal value, as given in \Cref{tab:Modelparameters}, in each panel.
    E-G) Heatmap plots of the colonisation metrics, $\tau_{tot}$ and $\tau_{max}$ for pairwise parameter variation.
    (E) and (F) present $\tau_{tot}$ and $\tau_{max}$ for varying $(U_B, \omega)$, respectively, and (G) presents $\tau_{tot}$ for varying ($U_B$, $Y_0$).
    Colour bars are given for each plot above the axis. 
    Grey points are used for the metric value for nominal parameters, yellow are times that exceed the nominal case, and blue are the times lower than the nominal case. 
    The rest of the metric data for pairwise parameter variation is presented in Supplementary Material \Cref{SM-SCATTER}. 
    }
    \label{fig:figure6}
\end{figure}

%% file: figures/figure7.tex
\begin{figure}[ht!]
\centering
    \includegraphics[width=\linewidth]{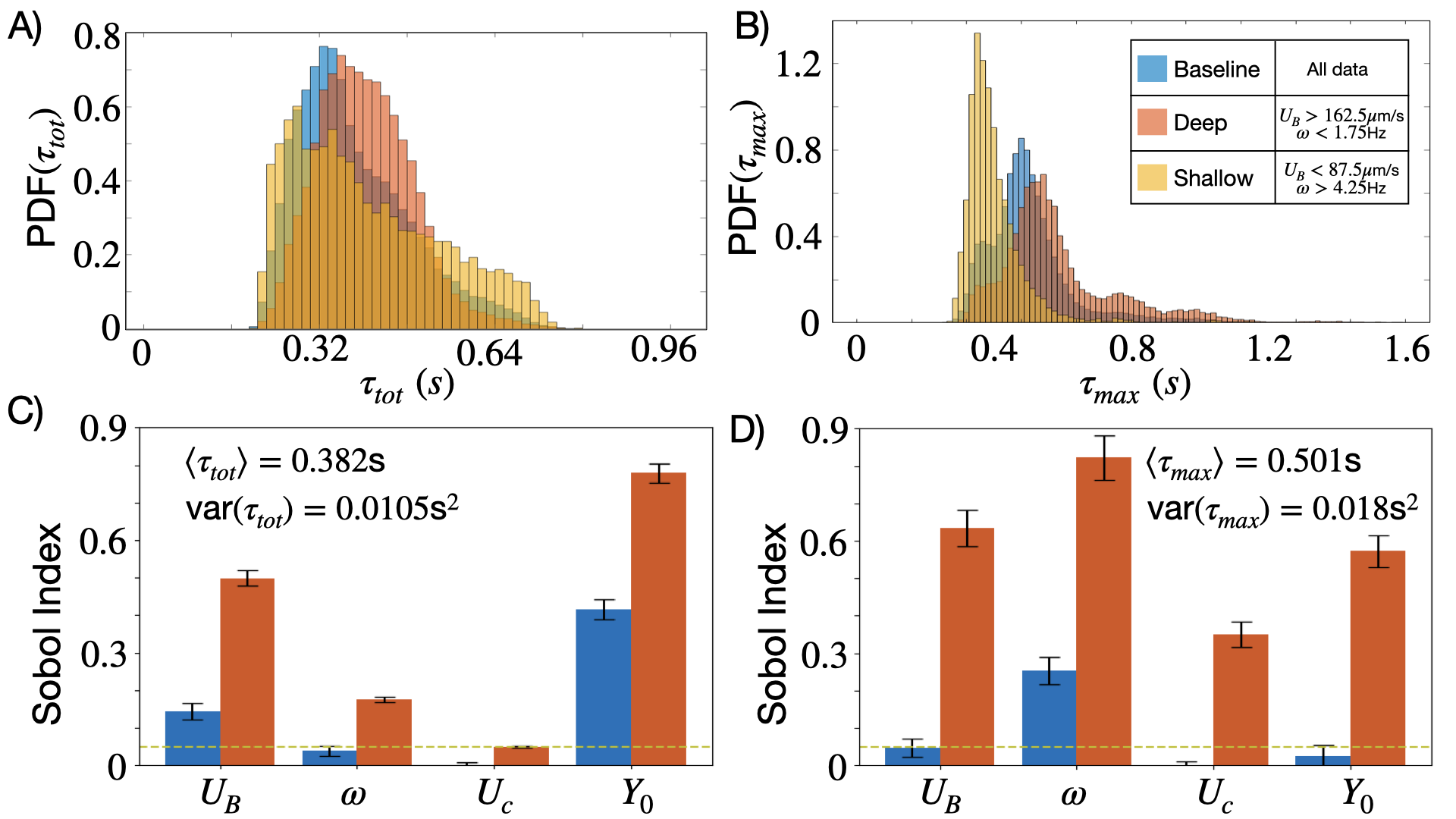}
    \caption{\textbf{Variance of Colonisation Metrics.} 
    A) Probability density function (PDF) of the average total time per bacterium spent in the critical zone, $\tau_{tot}$. 
    B) PDF of the maximum continuous time spent by a bacterium in the critical zone, $\tau_{max}$.
    For (A) and (B), three data sets are presented: baseline (blue), deep breathing (red), and shallow breathing (yellow).
    The legend shown in (B) provides the parameter bounds for the three sets.
    C) Sobol indices of the average total time spent in the critical zone, $\tau_{tot}$, for the baseline dataset including all simulations of varying parameters. 
    D) Sobol indices of the maximum continuous time spent in the critical zone, $\tau_{max}$, for the baseline dataset including all simulations of varying parameters. 
    For both (C) and (D), first-order Sobol indices (blue) and total-order Sobol indices (red) are shown, with error bars indicating the 95\% confidence interval.
    The 5\% significance is shown as a dashed line for comparison. 
    The mean and variance of the underlying dataset are presented in the inset.
    }
    \label{fig:figure7}
\end{figure}

%% file: discussion.tex
\section{Discussion}

We have presented a mathematical model to understand the colonisation of the Hawaiian Bobtail Squid, \textit{Euprymna scolopes}, by its luminous bacterial symbiont \textit{Vibrio fischerii}. 
The model simulates fluid flow and bacterial trajectories within the squid. 
We extend the model of \cite{nawroth_motile_2017} to include both the temporal dynamics of respiratory ventilation and confinement effects present \textit{in vivo}.
We focus on investigating the sensitivity of model parameters to elucidate the physiological responses of the squid to thermal stress.
Our data confirm that variations within a physically relevant parameter space can significantly affect the observed internal fluid flow, thereby dramatically reducing the time available for bacteria advected by these flows to be in proximity to the light organ.
In particular, parameter regimes of shallow breathing reduce the duration bacteria occupy space near the light organ.
Since this time is crucial for establishing symbiosis in the squid, such changes will compromise the squid's ability to form symbiotic relationships.
In turn, this will compromise their survivability in their native environment.

To quantify the impact of stress on the colonisation process, we focused our study on measuring the average total and maximum consecutive time intervals that bacteria advected by flows spent in a critical zone near squid appendages.
By using these two metrics, we account for bacteria colonisation from two characteristic motion patterns: first, those that, by the flow, are deposited onto the appendages, and second, those bacteria that spend long enough near the appendages to allow them, through chemotaxis, to transit the space.
Our results reveal strikingly distinct bacterial trajectories depending on the parameter regimes and corresponding conditions that give rise to them.
In general, some bacteria move along chaotic paths through the system, and others recirculate around the light organ appendages, directed by two vortices that alternately control the direction of the flow.
Our metrics indicate that the time spent in the colonisation region can be as low as 50\% of the nominal duration in certain regimes. 
The shortest observed durations are in systems with shallow breathing (\Cref{fig:figure7}B), which is expected when squid are under stress. 
\Cref{fig:figure6}E-G consistently show that these extremes in the parameter space lower the time bacteria can spend near appendages.
The Sobol indices (\Cref{fig:figure7}) further support the notion that changes in parameters can significantly impact the colonisation process. 
For $\tau_{max}$, the simultaneous varying of all of the parameters underlying the breathing dynamics can heavily impact the time.
In particular, this metric is significantly disrupted by univariate changes in breathing rate $\omega$, indicating that faster breathing, as expected during higher temperatures, can disrupt colonisation independently. 

Our results demonstrate that physiologically driven effects can significantly affect the duration of bacterial occupation near the light organ.
Although we observed some surprising univariate robustness, particularly in the maximum time bacteria spend in regions critical to colonisation, our results suggest that collective changes will reduce the time bacteria spend in these areas. 
This reduction indicates a potential vulnerability in the establishment process of symbiosis in squid, highlighting the susceptibility of this delicate mutualism to the cascading effects of thermal stress. 
In \cite{Otjacques_Climate_2025}, they note that at higher temperatures, a higher concentration of bacteria is required to combat reduced Es colonisation competency. 
Based on our simulation and this study, a higher concentration of Vf would allow more bacteria to achieve long maximum occupation times near the light organ, as well as increase the overall total time available for colonisation.
As such, our results provide one potential explanation for these observations, and further experimental studies are necessary to confirm our prediction that dynamic breathing changes drive this reduced competency.

Several factors in this model system remain to be explored in future work.
First, we neglect the swimming motility of the advected bacteria in our system, an acceptable approximation as the flow speeds present in our model, particularly in the vicinity of the light organ, far exceed those of Vf swimming.
Further modelling and computations would be needed to include bacterial motion.
Secondly, we explored a relatively large parameter space by conducting our study in two dimensions. 
We have sufficient data to perform a sensitivity analysis in a highly non-linear system.
Despite our progress, the squid internal cavity remains fundamentally three-dimensional.
Further experimental work \textit{in vitro} visualising temporally varying flows would help clarify the quality of these approximations.
Finally, we have taken the first steps toward incorporating confinement effects, leveraging several approximations to capture and understand the impact of mantle contraction and expansion.
We employ a time-dependent Poiseuille flow, and second, we utilise the trajectories of bacteria that were uniformly seeded along the inlet where this flow was imposed.
These assumptions are reasonable for the geometry of this system, but may not perfectly reflect the three-dimensional flow profile or initial distribution. 
Future work to address these issues and to incorporate squid physiology into the modelling is planned.

Understanding the dynamics driven by internal flows provides insight into vulnerabilities to a changing climate hidden within vital processes essential to a model system within symbiosis research, the Es-Vf system. 
Previous experimental results clearly show that stress drives changes in oxygen demand, which led us to choose the four breathing dynamics parameters we studied \cite{trubenbach_ventilation_2012,demont_effects_1984,melzner_temperature-dependent_2006,zielinski_temperature_2001}.
This study underscores the importance of considering physiologically driven physical processes when assessing ecological consequences, particularly in marine symbiotic interactions.
This modelling approach has enabled the exploration of a specific case of flow dynamics guided by ventilation and ciliary beating, phenomena common across organisms that regulate the movement of environmental bacteria.
Further research, expanding models and characterising breathing dynamics experimentally to validate the predictions presented here, will be crucial to understand the full extent of stress impacts and to explore potential mitigation strategies in light of global climate trends.

%% file: other.tex
\section*{Statements and Declarations}

\textbf{Conflict of Interest:} The authors declare that they have no known competing financial interests or personal relationships that could have appeared to influence the work reported in this paper.\\ \\
\textbf{Data Availability:} All code used in analysis during this study is included in the associated GitHub repository \cite{StephWilleniamsSquid}.\\ \\
\textbf{Use of Artificial Intelligence and AI-assisted technologies:} After the initial unassisted first draft preparation, Grammarly Pro was used to improve readability and language.

\section*{Acknowledgements}

The authors thank Ricardo Cortez for his input and discussions on the implementation of the Method of Regularised Stokeslets.

\section*{Funding} 
This material is based upon work supported by the National Science Foundation: DBI-2214038, and the Swedish Research Council under grant no. 2021-06594 while Shilpa Khatri was in residence at Institut Mittag-Leffler in Djursholm, Sweden, during the 2025 fall semester.



%% file: SM.tex
\section{Supplementary Materials}

\subsection{Stokes flow boundary conditions} \label{SM:BOUNDARY}

In this section, we provide details of the boundary conditions required to solve for the velocity and pressure when solving the Stokes equations, \Cref{StokesEquation_int} and \Cref{incompressibility}. 

There are three distinct types of boundaries in our model system: the external walls, an artificial boundary at the inlet to the internal cavity, and the appendage surfaces.
The boundary flow on the external walls (\Cref{fig:figure2}A: pink, blue, and green) is set to have a no-slip boundary condition, $\bm{u}(\bm{X}(\ell),t) = 0$. 

A Poiseuille flow is enforced on the artificial inlet boundary of the wide opening of the internal cavity (\Cref{fig:figure2}A: peach). The magnitude is given by a time-dependent function, acting in the $\hat{y}$ direction, 
\begin{equation}
    u_2(\bm{X}(\ell),t) = \frac{1}{2}U_B\left( 1 + \cos( 2 \pi \omega t \right) \left( 1 - \left(\frac{2|r|}{W_B}\right)^2 \right) \, ,
    \label{timeDependenceVent}
\end{equation}
where $\bm{u}(\bm{x},t) = (u_1(\bm{x},t),u_2(\bm{x},t))$, $\bm{X}(\ell)$ is on the inlet boundary, $|r|$ is the distance of $\bm{X}(\ell)$ from the centre bisecting the channel (see Figure 2A; grey line from $W_B$ to $W_T$), $W_B$ is the width of the opening, $\omega$ is the frequency of the flow, and $U_B$ is the maximum magnitude of the flow. 
The flow in the $\hat{x}$ direction is set such that $u_1(\bm{X}(\ell),t) = 0$. 
The temporally varying magnitude of this flow enables us to model the squid's breathing dynamics.

Finally, the boundary conditions on the appendage surfaces (\Cref{fig:figure2}B) is given by the function,

\begin{equation}
    \bm{u}_s(\psi) = 
    \begin{cases}
    U_{C} \sin \left( \frac{4}{3} (\psi - \psi_0) \right) \bm{\hat\psi}, & 2\pi N + \psi_0 \leq \psi \leq 2\pi N + \frac{3\pi}{2} + \psi_0, ~\text{with } N \in \mathbb{Z} \, ,\\
    0 \, , & \text{otherwise} \, .
    \end{cases}
    \label{boundaryflow}
\end{equation}
The appendage surface is parametrised by the angle $\psi$, so that  $u(\bm{X}(\ell),t) =\bm{u}_s(\psi)$, when $\bm{X}(\ell)$ is on the surface of the appendages. 
Here, $\psi_0$ is used to orient each appendage, and $U_C$ is the maximum time-averaged flow speed of the cilia beating. 
Note that the sign of $\bm{u}_s(\psi)$ can be positive or negative, modelling the changing direction of cilia beating. 
The appendage radius, $R$, encompasses the appendage itself, as well as the length of the ciliated hairs. The flow, $\bm{u}_s(\psi)$, is prescribed at $R_{\text{eff}}=0.9R$, chosen to be interior to this radius by half the length of the cilia \cite{nawroth_motile_2017}, see \Cref{fig:figure2}B. 

By using these prescribed boundary conditions, we can solve for the resulting fluid flow within the squid as it breathes.

\subsection{The Method of Regularised Stokeslets} \label{SM-MRS}

An overview of the numerical methods to solve the mathematical model is presented in \Cref{NumericalMethods}. 
Here, we provide further details of the Method of Regularised Stokeslets. 

The Stokes equations can be solved using the Green's function, the Stokeslet. 
However, due to the resulting Stokeslet singularity, these point-force solutions pose challenges when using numerical methods. 
Instead, we use the Method of Regularised Stokeslets to solve the momentum equation, Equation \eqref{StokesEquation} with the divergence-free constraint, \Cref{incompressibility}, as discussed in \Cref{NumericalMethods} \cite{cortez_method_2001, cortez_method_2005}. 

In this method, a regularised Stokeslet is solved for using a regularised delta function force, also known as a blob function. 
We choose the following commonly used blob in two dimensions,

\begin{equation}
    \phi_\epsilon(\bm{x}) = \frac{3\epsilon^3}{2\pi(||\bm{x}||^2 + \epsilon^2)^{5/2}}, 
\end{equation}
resulting in the regularised Stokeslet tensor, 
\begin{equation}
S_{mn}^{\epsilon}(\bm{x},\bm{x}_i) = \frac{1}{4\pi\mu} \left(-\log(R_{\epsilon}) \delta_{mn} + \frac{(x_m - x_{im})(x_n - x_{in})}{ {R_{\epsilon}}^2 } \right),
\end{equation}
where
\begin{eqnarray}
R_{\epsilon} = \sqrt{||\bm{x}-\bm{x_i}||^2 + \epsilon^2} + \epsilon \, , 
\end{eqnarray}
$\bm{x} = (x_1,x_2)$, $\bm{x}_i = (x_{i1},x_{i2})$, and $\delta_{mn}$ is the Kronecker delta. 

To solve for the forces at a time $t$, $\bm{f}_i(t)$ on the boundary points, $\bm{U}_0(t)$, and $\Omega(t)$, required to calculate the flows in the system, we construct a linear system by setting $\bm{x} = \bm{x_j}$, $j = 1 \dots N$ in \Cref{flow_calculation}, and include the net-zero force and torque constraints with these equations,

\begin{equation}
    \begin{bmatrix}
        \bm{S}_{11}^\epsilon & \bm{S}_{12}^\epsilon & \hdots & \bm{S}_{1N}^\epsilon & -\bm{I}_2 & -\bm{K}_1\\
        \bm{S}_{21}^\epsilon & \bm{S}_{22}^\epsilon & \hdots & \bm{S}_{2N}^\epsilon & -\bm{I}_2 & -\bm{K}_2\\
        \vdots & \vdots & \hdots & \vdots \\
        \bm{S}_{N1}^\epsilon & \bm{S}_{N2}^\epsilon & \hdots & \bm{S}_{NN}^\epsilon & -\bm{I}_2 & -\bm{K}_N\\
        \bm{I}_2 & \bm{I}_2 & \hdots & \bm{I}_2 & \bm{0}_{22} & \bm{0}_{21}\\
        \bm{K}_1^T & \bm{K}_2^T & \hdots & \bm{K}_N^T & \bm{0}_{12} & 0\\
    \end{bmatrix}
    \begin{bmatrix}
        \bm{f}_1\\
        \bm{f}_2\\
        \vdots\\
        \bm{f}_N\\
        \bm{U}_0\\
        \Omega
    \end{bmatrix} 
    =
    \begin{bmatrix}
        \bm{u}_1\\
        \bm{u}_2\\
        \vdots\\
        \bm{u}_N\\
        \bm{0}_{21}\\
        0
    \end{bmatrix}.
    \label{total_stokeslet}
\end{equation}
Here, $\bm{S}_{ji} = S^{\epsilon}(\bm{x}_j,\bm{x}_i)$ is the regularised Stokeslet tensor between the force at $\bm{x}_i$ acting on the position $\bm{x}_j$. 
$\bm{I}_2$ is the $2\times2$ identity matrix, and $\bm{0}_{pq}$ is an $p \times q$ matrix of zeros.
The values of $\bm{u}_j(t) = \bm{u}(\bm{x}_j,t)$ come from the boundary conditions of the system, as discussed in \Cref{SM:BOUNDARY}.
Since each of the Stokeslets in this formulation depends only on the geometry of the boundaries (and not on the time-varying boundary conditions), and the geometry is assumed static, this matrix is time-independent.
As such, the matrix and its inverse need only to be calculated once, and then the inverse can be applied to any set of boundary conditions to give the set of forces necessary to satisfy these conditions. 
Once we have the forces at a specific time, we can then compute the velocity at that time at any point $\bm{x}$ in the squid domain using \Cref{flow_calculation}.

The Method of Regularised Stokeslets allows us to compute an approximate solution to the Stokes equations, \Cref{StokesEquation_int} and \Cref{incompressibility}.
To verify the method and implementation, we compute velocity solutions on a uniform grid with boundary conditions set at $t = 0$ (as shown in \Cref{fig:figure3}B) for increasing $\rho$ and proportionally decreasing $\epsilon$.   
The grid is of size $1260 \times 1710$ $\mu$m with $280 \times 380$ points uniformly spaced to cover the entire internal cavity.  
We compute the $L_2$  and $L_\infty$ error of the velocity on the points on the grid for consecutive densities, $\rho$. 
We find second-order convergence in $\epsilon$, as expected. 
We choose a density of $\rho = 30$ where the $L_2$ error in the y-component of the fluid velocity for our test case, for consecutive densities, is approximately 1.10\% and the $L_\infty$ error in the velocity for consecutive densities is approximately 1.59\%.

\subsection{Parameter ranges} \label{SM-PARAMS}

As discussed in \Cref{Metrics}, we choose to vary the system parameters most affected by thermal stresses. 
The four varied parameters are the ventilation flow strength, $U_B$, the breathing frequency, $\omega$, the maximum cilia flow strength, $U_C$, and the initial distance of the bacteria from the appendages, $Y_0$.

We calculate the fluid flux through the system during a breath cycle to determine the range of values to use for the breathing strength, $U_B$.
Assuming the inlet is cylindrical and the fluid flow is a Poiseuille flow as prescribed by the boundary condition, \Cref{timeDependenceVent}, the volume of fluid passing through a cross-section of the inlet in three dimensions during one breath cycle is,
\begin{equation}
V = \int_0^{2\pi/\omega} \int_{0}^{W_B/2} \int_{0}^{2\pi} \frac{U_B}{2} (1 + \cos(\omega t)) \left( 1 - \left(\frac{2r}{W_B}\right)^2 \right) r \, d\theta dr dt = \frac{\pi^2 U_B W_B^2}{8 \omega} \, . 
\label{volume_circular}
\end{equation}
The volume of fluid passing through a cross-section of the outlet nozzle can be similarly computed. 
In \cite{nawroth_motile_2017}, they measured an outlet flow speed of $U_T = 500 ~\mu\text{m s}^{-1}$. 
Therefore, when $U_T=500$ $\mu\text{m s}^{-1}$, $W_T = 400$  $\mu$m, and $\omega=3$, the fluid flux is 0.033 $\mu$l per breath cycle. 
Since the volume flux for the inlet and outlet must be the same, with an inlet width of $U_B = 1000$ $\mu\text{m }$, the ventilation flow speed is $80~\mu\text{m s}^{-1}$, close to the proposed speed of $100~\mu\text{m s}^{-1}$ in \cite{nawroth_motile_2017}.

Since we are working in two dimensions,  the fluid flux through a cross-section of the inlet is,

\begin{equation}
V = \int_0^{2\pi/\omega} \int_{0}^{W_B/2} \frac{U_B}{2} (1 + \cos(\omega t)) \left( 1 - \left(\frac{2r}{W_B}\right)^2 \right) dr dt = \frac{\pi U_B W_B}{3 \omega} \, . 
\label{volume_pipe}
\end{equation}
In this case, when the outlet flow speed is $U_T = 500 ~\mu\text{m s}^{-1}$, the ventilation flow speed is $U_B = 200 ~\mu\text{m s}^{-1}$. 

To span both the two and three dimensions cases, we choose the range of values for $U_B$ to be $50 ~\mu\text{m s}^{-1}$ to $200~\mu\text{m s}^{-1}$ with a nominal value of $125~\mu\text{m s}^{-1}$ close to the proposed value of \cite{nawroth_motile_2017}.

Next, the range for $\omega$ is chosen based on measurements presented in \cite{visick_exclusive_2000}.
This paper proposes an approximate range of 3 to 4 breaths per second as nominal.
To sample a physiological range, we use a range of 0.5 to 5.5 breaths per second.

The value of nominal ciliary-driven flow is from \cite{nawroth_motile_2017}, set at $600~\mu\text {m s}^{-1}$.
We select the range around this to be $\pm 200~\mu\text{m s}^{-1}$. 

By varying the location at which we seed our particles upstream of the appendages, $Y_0$, we are also able to account for at what point in the breath cycle the particle arrives at the inlet.
This also allows us to indirectly account for squid of different sizes and, therefore, various ages.
We set the range of $Y_0$ between 135$~\mu$m and 315$~\mu$m upstream of the appendage centres.

\subsection{Data scatter plots} \label{SM-SCATTER}

\begin{figure}[ht!]
\centering
    \includegraphics[width=1\linewidth]{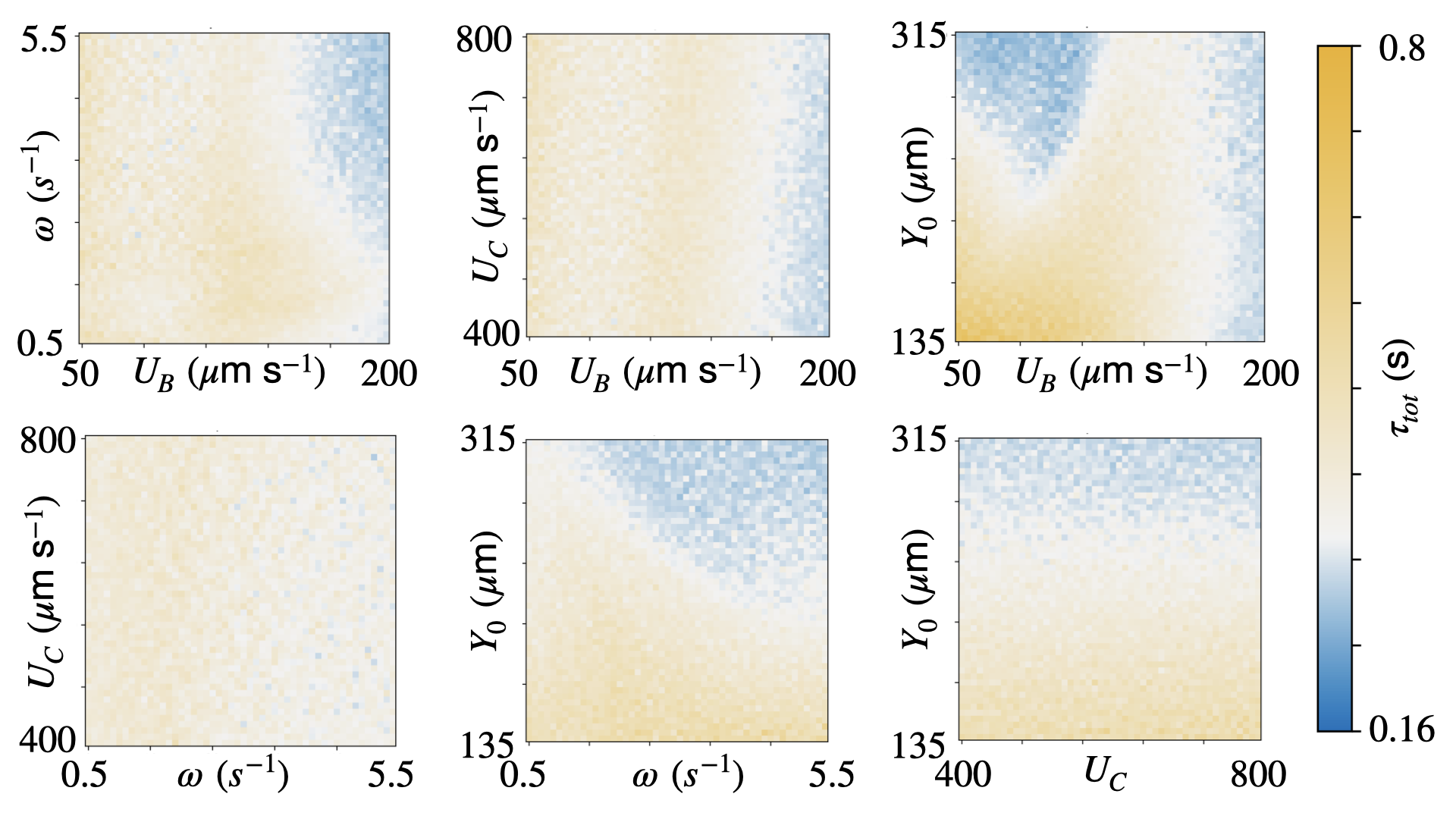}
    \caption{\textbf{Heatmap plots of the colonisation metric, $\tau_{tot}$.} 
    Heatmaps of $\tau_{tot}$ are presented for varying parameter pairs. 
    The colour bar is scaled so that the metric for the nominal parameter values is grey. Blue bins represent regions with a metric value lower than the nominal parameters, and the yellow bins represent regions with a higher metric value.
    }
    \label{fig:TtotHeatmap}
\end{figure}

\begin{figure}[ht!]
\centering
    \includegraphics[width=\linewidth]{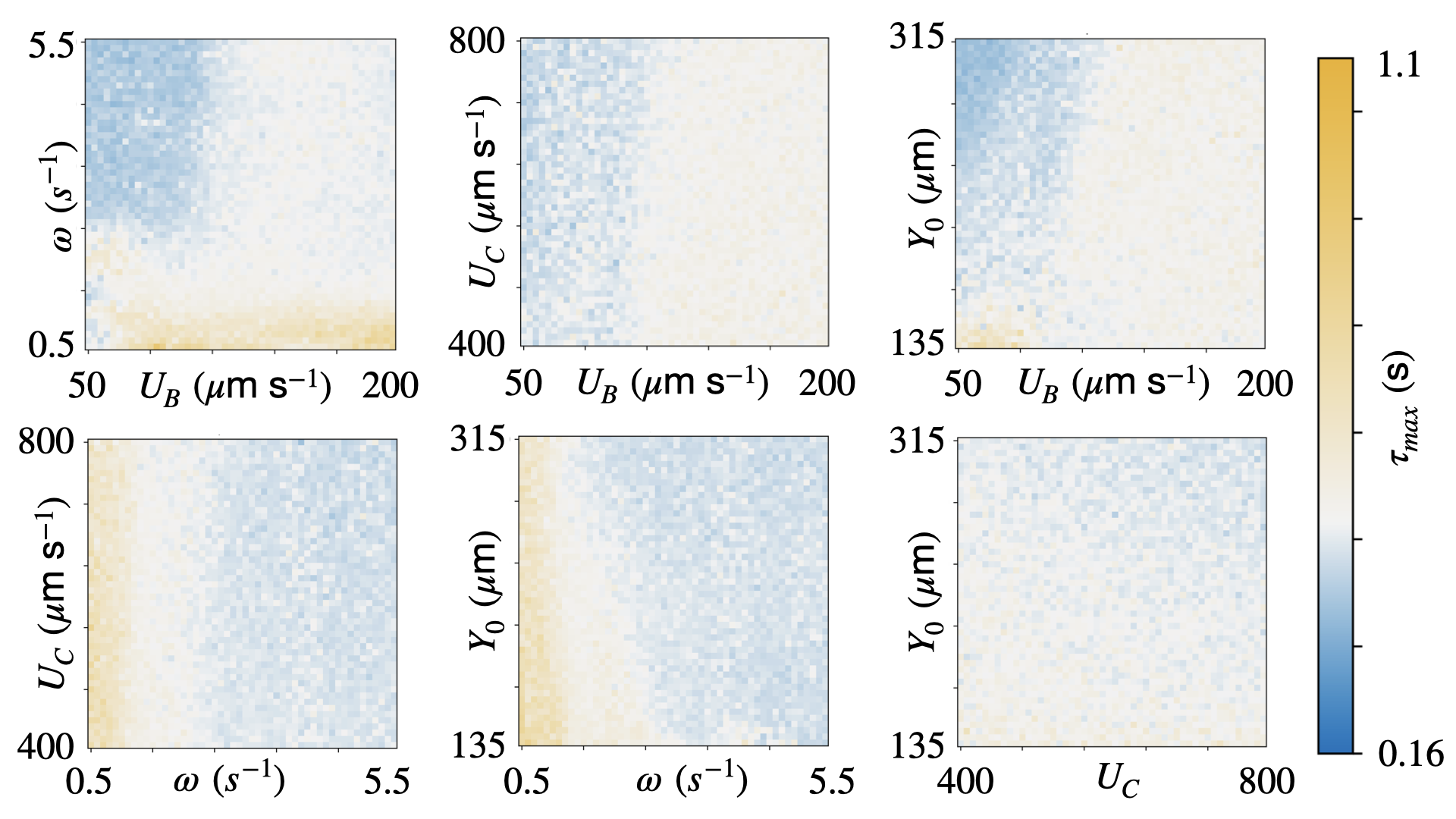}
     \caption{\textbf{Heatmap plots of the colonisation metric, $\tau_{max}$.} 
    Heatmaps of $\tau_{max}$ are presented for varying parameter pairs. 
    The colour bar is scaled so that the metric for the nominal parameter values is grey. Blue bins represent regions with a metric value lower than the nominal parameters, and the yellow bins represent regions with a higher metric value.}
    \label{fig:TmaxHeatmap}
\end{figure}

As discussed in \Cref{sec:extremes}, all of the heatmap plots of the colonisation metrics, $\tau_{tot}$ and $\tau_{max}$, for pairwise parameter variation are presented in \Cref{fig:TtotHeatmap} and \Cref{fig:TmaxHeatmap}, respectively.
We bin the data according to the range of the underlying parameters. 
Each parameter range is split into 50 uniform bins, resulting in a 50x50 grid for each heatmap. 
In each bin, we calculate and present the mean of the metric. 

In all of the heatmaps, yellow and blue bins are present, mixed throughout the phase space. The same is true of the underlying data prior to averaging.
In regions where one colour is dominant (i.e., the bacteria spend more or less time than in the nominal case), the remaining two parameters not plotted can counter local trends.
This resulting mixture of bins indicates that the omitted parameters in each plot can overpower local trends.
Throughout the data presented, we observe that the metrics are nonlinearly dependent on the parameters. 

\subsection{Sobol second order indices} \label{SM-SOBOL2}

\begin{figure}[ht!]
\centering
    \includegraphics[width=0.8\linewidth]{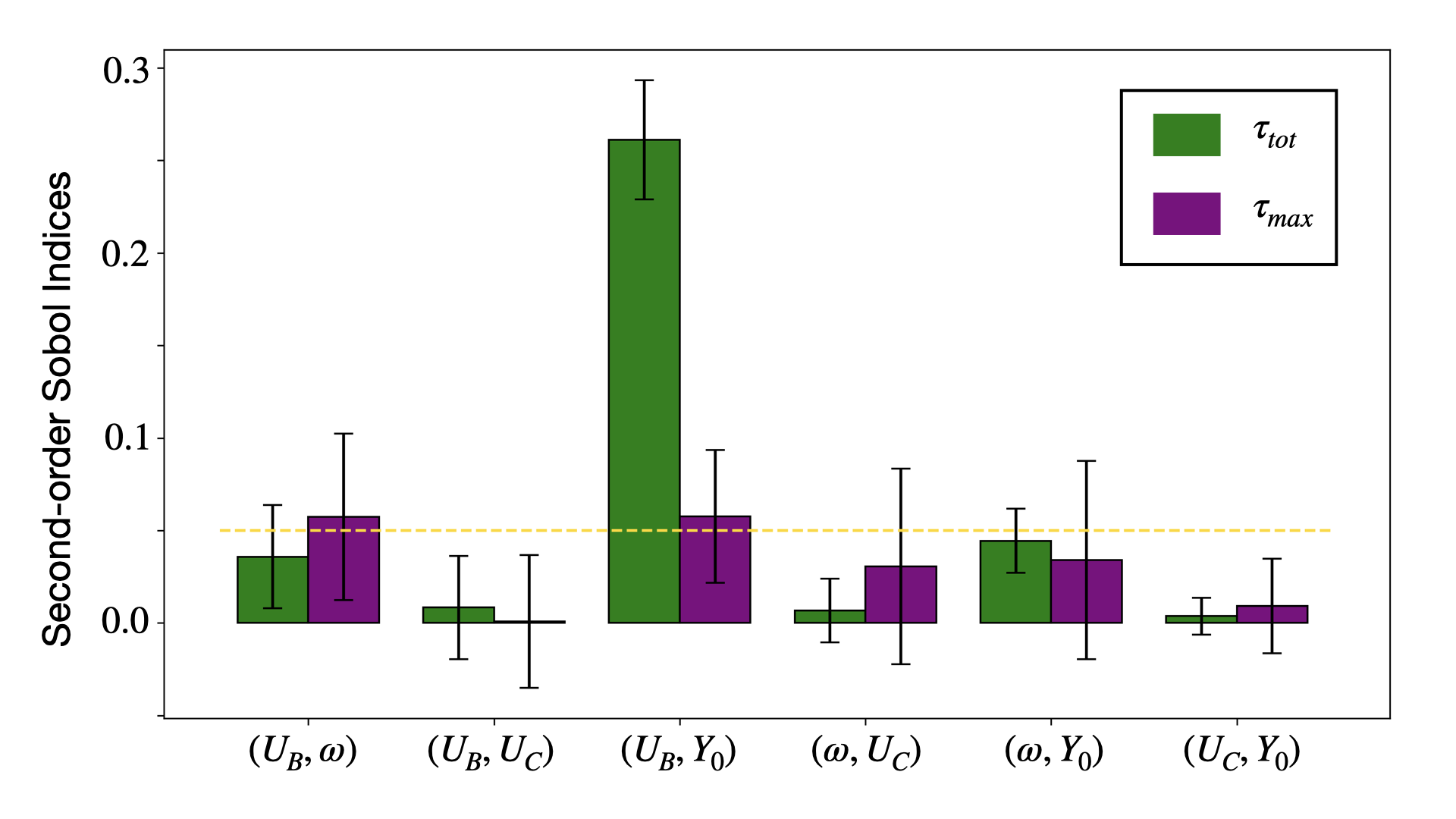}
    \caption{\textbf{Second-order Sobol indices of the colonisation metrics.} 
    The second-order Sobol indices for $\tau_{tot}$ (green) and $\tau_{max}$ (purple) are shown for the baseline dataset, including all simulations with varying parameters. 
    Error bars indicate 95\% confidence intervals.
    The 5\% significance threshold is shown as a dashed line for comparison. 
    The corresponding first and total order indices are presented in \Cref{fig:figure7}. 
    }
    \label{fig:Sobol2}
\end{figure}

The values of the second-order Sobol indices for $\tau_{tot}$ and $\tau_{max}$ are given in \Cref{fig:Sobol2}, corresponding to the first and total order indices discussed in \Cref{sec:importance}.
The 95\% confidence intervals are given as error bars, calculated by bootstrapping in SALib \cite{Iwanaga_Toward_SALib_2_0_2022}.
Most of the second-order interactions make a very minor contribution to the variance in the metrics.
The interaction $(U_B,Y_0)$ is the only exception for $\tau_{tot}$. 
$(U_B,\omega)$ and $(U_B,Y_0)$ are very slightly above the 5\% threshold for $\tau_{max}$.
All three of these pairings suggest that variations in ventilation strength, $U_B$, can interact with variations in the other parameters to drive larger deviations from the nominal dynamics and resulting metrics.

